\documentclass[prc,notitlepage,showkeys,11pt,a4paper,tightenlines]{revtex4}
\usepackage{amsmath}
\usepackage{amssymb}
\usepackage{graphicx}
\usepackage{booktabs}
\usepackage[usenames, dvipsnames]{color}
\usepackage[caption=false]{subfig}

\begin{document}

\title{Statistical analysis of beta decays and the effective value of $\boldsymbol{g}_\text{A}$ in the\\[1mm]proton-neutron quasiparticle random-phase approximation framework}

\author{Frank F. \surname{Deppisch}}
\email{f.deppisch@ucl.ac.uk}
\affiliation{Department of Physics \& Astronomy, University College London,\\London WC1E 6BT, United Kingdom}

\author{Jouni \surname{Suhonen}}
\email{jouni.suhonen@phys.jyu.fi}
\affiliation{Department of Physics, University of Jyvaskyla, P.O. Box 35,
FI-40014 Jyvaskyla, Finland}

\begin{abstract}
We perform a Markov Chain Monte Carlo (MCMC) statistical analysis of a number of 
measured ground-state-to-ground-state single $\beta^+$/electron-capture and $\beta^-$ 
decays in the nuclear mass range $A=62-142$. The corresponding experimental 
comparative half-lives ($\log ft$ values) are compared with the theoretical ones 
obtained by the use of the proton-neutron quasiparticle random-phase approximation 
(pnQRPA) with G-matrix based effective interactions. The MCMC analysis is performed 
separately for 47 isobaric triplets and 28 more extended isobaric chains of nuclei 
to extract values and uncertainties for the effective axial-vector coupling 
constant $g_\text{A}$ in nuclear-structure calculations performed in the pnQRPA 
framework. As far as available, measured half-lives for two-neutrino double 
beta-minus decays occurring in the studied isobaric chains are analyzed as well.
\end{abstract}


\maketitle


\section{Introduction}

The neutrinoless double beta ($0\nu\beta\beta$) decays of atomic nuclei serve as a 
forceful incentive to constantly drive nuclear-structure calculations toward better 
performance. Analyses of the potential experimental $0\nu\beta\beta$ outcomes in the 
future require accurate knowledge of the related nuclear matrix elements (NMEs) in 
order the obtained data to serve in best possible ways to unravel the fundamental 
nature and mass of the neutrino 
\cite{REPORT, Vergados2012, Deppisch:2012nb, Rodejohann:2011mu}. It is also tightly 
connected to the breaking of lepton number asymmetry and has far reaching consequences 
even on solutions on the baryon asymmetry of the Universe 
\cite{Deppisch:2015yqa, Deppisch:2004kn}. A host of models, ranging from the 
interacting shell model (ISM) to various mean field theories, have been used in the 
calculations. The resulting NMEs have been analyzed in the review article 
\cite{Suhonen2012d}. Most of the calculations have been pursued in the framework 
of the proton-neutron quasiparticle random-phase approximation (pnQRPA) \cite{Suhonen2012c}.

In these many calculations it has been noticed that several aspects of nuclear 
structure make an impact on the resulting values of the NMEs: the chosen valence 
space and orbital occupancies \cite{Suhonen2008b, Suhonen2010, Suhonen2011b}, the 
shell-closure effects \cite{Suhonen2012d, Barea2009} and the deformation 
\cite{Alvarez2004, Caurier2008b, Menendez2009a, Rodriguez2010}. Only lately the 
important aspect of the effective value of the axial-vector coupling constant 
$g_\text{A}$ has been addressed within few models like the pnQRPA 
\cite{Faessler2008, Suhonen2013c, Holt2013, Suhonen2014, Delion2014}, the Interacting Shell Model (ISM) \cite{Wildenthal1983, Martinez-Pinedo1996, Menendez2011, Caurier2012} and 
the Interacting Boson Model 2 (IBA-2) \cite{Barea2013}.

A particular problem with the pnQRPA calculations, not present in the other 
calculations, is the unsettled value of the particle-particle interaction parameter 
$g_\text{pp}$ describing the strength of the proton-neutron interaction in the $1^+$ 
channel. Since the introduction of this parameter \cite{Vogel1986, Civitarese1987} 
its values have been tried to fix by the inspection of the measured single-beta-decay 
rates \cite{Suhonen2005, Suhonen2011d} or $2\nu\beta\beta$-decay rates 
\cite{Rodin2006, Kortelainen2007b, Kortelainen2007c, Suhonen2008a}.

Here we make an attempt to relate the values of $g_\text{pp}$ to the values of 
$g_\text{A}$ through the data on beta-decay rates associated with the transitions 
between an even-even and an odd-odd nucleus. The data on these decays are presented 
as comparative half-lives ($\log ft$ values) and comparing them with the corresponding 
computed ones one can make conclusions about the possible correlations of these two 
key parameters of calculation. As mathematical aid we use the Markov Chain Monte 
Carlo (MCMC) statistical analysis of 47 isobaric triplets and 28 more extended 
isobaric chains of nuclei. In the isobaric triplets there are two beta-decay branches, 
left and right, between the central and lateral nuclei, and in the extended isobaric 
chains more complex systems of consecutive central and lateral nuclei can form. 
To estimate the theoretical uncertainty inherent in the pnQRPA framework 
we include the full parametric freedom available. This means we introduce an uncertainty 
in the particle-hole interaction parameter $g_\text{ph}$. In addition we treat 
both $g_\text{pp}$ and $g_\text{ph}$ as parameters specific only to a given 
$\beta^+$/electron capture (EC) or $\beta^-$ decay transition pair. This opens up a 
large parametric freedom that has not been explored before.

Our analysis is intended to address the importance of quenching, i.e. the 
suppression of $g_\text{A}$ with respect to its free value $g_\text{A} = 1.269$. 
Quenched values as low as $g_\text{A} \approx 0.4$ have been reported for example 
in the IBM-2 model \cite{Barea2013}; because $0\nu\beta\beta$ decay depends on 
$g_\text{A}$ as $\propto g_\text{A}^4$, this could reduce the decay by orders of 
magnitude, having a serious impact on the observability of $0\nu\beta\beta$ decay 
in experiments. Whether such strong quenching actually applies to $0\nu\beta\beta$ is 
not a question we can answer here, because we touch here on only the $1^+$ multipolarity of
the multipole decomposition of a $0\nu\beta\beta$ decay NME and because the
$0\nu\beta\beta$ decay proceeds via a momentum exchange much larger than that of the 
presently discussed single and two-neutrino double beta decays.

The article is organized as follows. In Sec.~\ref{sec:theory} the basic theoretical 
framework is briefly reviewed and the model-space aspects and the adjustment of the model 
parameters are explained. In Sec.~\ref{sec:quenching} we make a statistical analysis 
of the effective value of $g_{\rm A}$. Here we try to chart the possible effective 
values of the axial-vector coupling constant $g_\text{A}$ in model calculations using 
the pnQRPA approach, using different methods. Finally, in Sec.~\ref{sec:conclusions}, 
we summarize and draw the conclusions.


\section{Brief summary of the theory}
\label{sec:theory}

We begin by defining the comparative half-lives ($\log ft$ values) of the 
$1^+\leftrightarrow 0^+$ Gamow--Teller transitions that form the basis of the present 
analysis. The $\log ft$ value is defined as \cite{Suhonen2007}
\begin{align} 
\label{eq:logft}
	\log ft = \log_{10} (f_0 t_{1/2}[s]) = \log_{10} \left(\frac{6147}{B_\text{GT}}\right),
\end{align}
with
\begin{align} 
\label{eq:nmeGT}
	B_\text{GT} = 
	\frac{g_\text{A}^2}{2J_i+1}\left\vert M_\text{GT}(g_\text{pp}, g_\text{ph}) \right\vert^2
\end{align}
for the $\beta^+$/EC or $\beta^-$ type of transitions. Here the 
half-life $t_{1/2}$ has been given in seconds and $f_0$ is the dimensionless leptonic phase space factor associated with the process. $J_i$ is the spin of the initial 
ground state and $M_\text{GT}$ is the Gamow-Teller NME defined, e.g. in Ref.~\cite{Suhonen2007}. Here $g_\text{A}$ is the weak axial-vector coupling constant, 
$g_\text{pp}$ the particle-particle interaction coupling constant and $g_\text{ph}$ 
the particle-hole interaction coupling constant as defined, e.g. in Refs.~\cite{Suhonen1988a, Suhonen1988b}. Methods of determining the values of these 
constants are addressed in  the next section.

The $2\nu\beta\beta$ decay half-life can be compactly written as
\begin{equation}
\label{eq:2vbb}
	\left[ t_{1/2}^{(2\nu)}(0_i^+ \to 0_f^+) \right]^{-1} = 
	g_\text{A}^4 G_{2\nu} \left\vert M^{(2\nu)}\right\vert^2,
\end{equation}
where $G_{2\nu}$ stands for the leptonic phase-space factor without including 
$g_\text{A}$ in the way defined in Ref.~\cite{Kotila2012}. The initial ground state is 
denoted by $0^+_i$ and the final ground state by $0^+_f$. The $2\nu\beta\beta$ 
NME $M^{(2\nu)}$ can be written as
\begin{align}
\label{eq:NMEb}
	M^{(2\nu )} = \sum_{m,n} 
	\frac{M_\text{F}(1^+_m) \langle 1^+_m\vert 1^+_n\rangle M_\text{I}(1^+_n)}{D_m},
\end{align}
where the quantity $D_m$ is the energy denominator containing the average energy of 
the $1^+$ states emerging from the pnQRPA calculations in the initial and final 
even-even nuclei. The summation is in general over all intermediate $1^+$ states 
where $\langle 1^+_m\vert 1^+_n\rangle$ is the overlap between two such states. 
We will in general treat the individual matrix elements for the transition between 
the initial (final) state and the virtual intermediate states, $M_\text{I(F)}(1^+_i)$ 
as functions of separate sets of $g_\text{pp}$ and $g_\text{ph}$ couplings. 

While we do not discuss $0\nu\beta\beta$ decay in detail in this article, we 
briefly describe the theoretical calculation of the corresponding half-life to 
illustrate the similarities and differences to the above processes. The $0\nu\beta\beta$ decay half-life can be written as
\begin{equation}
\label{eq:0vbb}
	\left[ t_{1/2}^{(0\nu)}(0_i^+ \to 0_f^+) \right]^{-1} = 
	g_{\text{A},0\nu}^4 
	\left(\frac{\langle m_\nu \rangle}{m_e}\right)^2  
	G_{0\nu} \left\vert M^{(0\nu)}\right\vert^2,
\end{equation}
where $G_{0\nu}$ stands for the leptonic phase-space factor without including 
$g_{\text{A},0\nu}$ and the electron mass $m_e$ in the way defined in Ref.~\cite{Kotila2012}. 
Here, we denote the effective axial coupling relevant for $0\nu\beta\beta$ decay as
$g_{\text{A},0\nu}$ to emphasise that its value may deviate from the one determined 
in single beta and $2\nu\beta\beta$ decays. The effective $0\nu\beta\beta$ neutrino 
mass is denoted as $\langle m_\nu \rangle$. As before, the initial ground state is 
denoted by $0^+_i$ and the final ground state by $0^+_f$. 

The $2\nu\beta\beta$ decay and $0\nu\beta\beta$ decay half-lives share the same 
strong dependence on $g_{\text{A}}$ as seen in Eqs.~(\ref{eq:2vbb}) and (\ref{eq:0vbb}).
It is thus an essential first step to study the effective value of $g_{\text{A}}$ in single
beta and $2\nu\beta\beta$ decays. These studies tangent only the $1^+$ contribution to
the $0\nu\beta\beta$ decay whereas it is known that higher multipoles are very
important for the $0\nu\beta\beta$ decay as well \cite{Hyvarinen2015}. Some attempts
to study these higher multipolarities by way of single beta decays have been
made lately \cite{Ejiri2014, Haaranen2016}. It is thus not straightforward
to relate the single and $2\nu\beta\beta$ decay studies to the value of the 
$0\nu\beta\beta$ NME, especially since the former involve momentum transfers of
a few MeV and the latter involves a virtual neutrino with a momentum exchange
of the order of 100~MeV. This allows the possibility that the effective value of $g_{\text{A}}$ 
gets momentum dependent \cite{Menendez2011}. Related to this, the high momentum exchange
in $0\nu\beta\beta$ decay makes the higher $J^{\pi}$ states contribute appreciably
to the decay rate \cite{Hyvarinen2016}. For these higher-lying states the quenching
of $g_{\text{A}}$ could be different from the low-lying states discussed in the
present work. 
It should be noted, however, that in the pnQRPA no closure approximation is imposed in either modes of double beta decay so that the individual contribution from all intermediate states can be accessed in the case of $0\nu\beta\beta$ decay, as well. These intermediate contributions vary strongly from nucleus to nucleus and even some kind of single-state dominance can be observed for some $0\nu\beta\beta$ decaying nuclei \cite{Hyvarinen2016}. For more details on the theoretical background, we refer the reader to Ref.~\cite{Suhonen2014}.

\begin{figure}[t!]
\centering
\includegraphics[width=0.8\columnwidth]{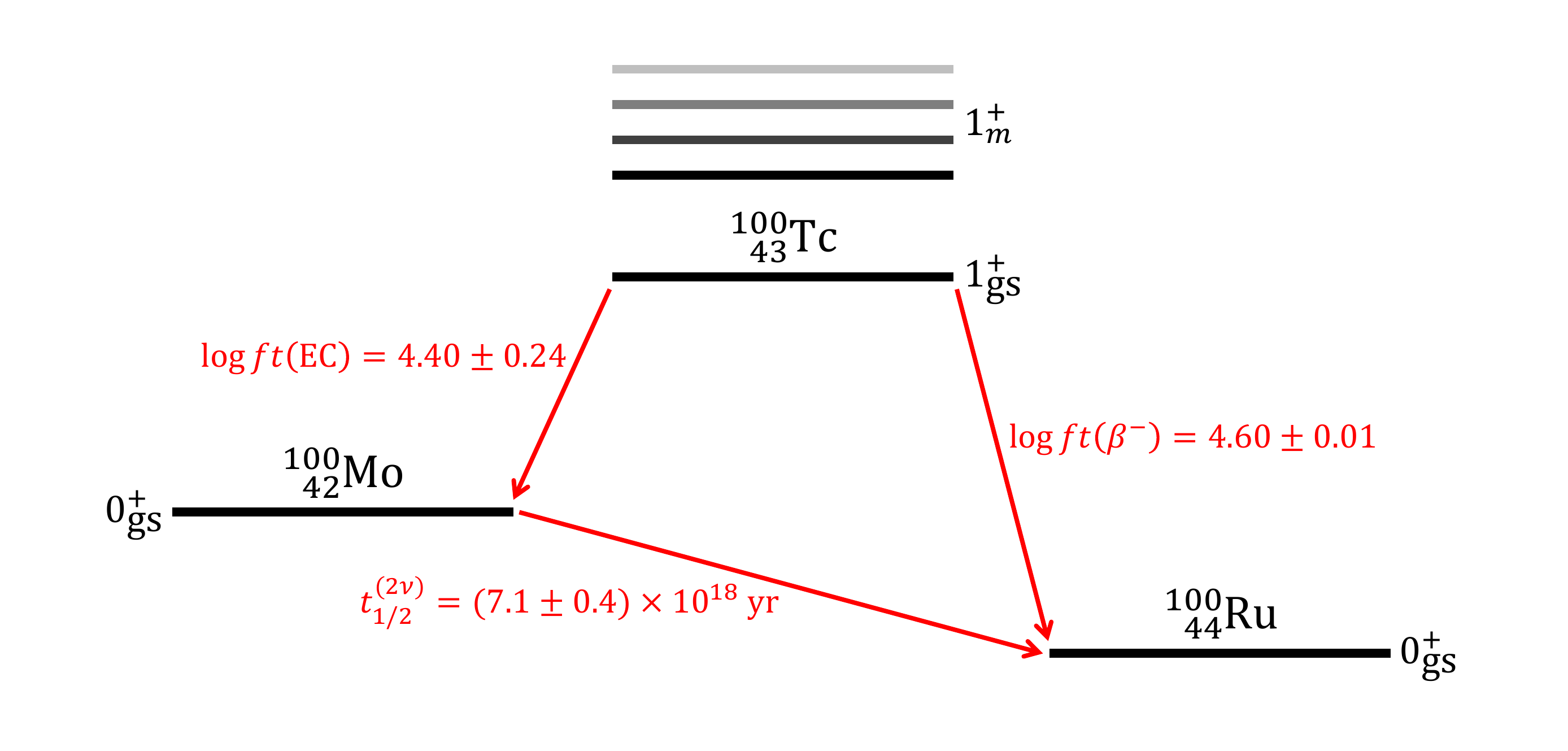}
\caption{Double and single beta decay characteristics of the isobaric triplet 
$^{100}_{\ 42}\text{Mo}$, $^{100}_{\ 43}\text{Tc}$ and $^{100}_{\ 44}\text{Ru}$. 
The experimental $2\nu\beta\beta$ half-lives and the $\log ft$ values are discussed 
in Sec.~\ref{sec:quenching}.}
\label{fig:schemeMo}
\end{figure}
In the present calculations we obtain the single-particle energies from a spherical 
Coulomb-corrected Woods--Saxon (WS) potential with the standard parametrization of 
Bohr and Mottelson \cite{Bohr1969}. This parametri\-zation is optimized for nuclei 
near the line of beta stability and is thus well suited for the presently studied 
nuclei. The single-particle orbitals used in the calculations span the space 
0f-1p-0g-2s-1d-0h$_{11/2}$ for the masses $A=62-80$, 0f-1p-0g-2s-1d-0h for the 
masses $A=98-108$ and 0f-1p-0g-2s-1d-0h-1f-2p for the masses $A=110-142$. In these
single-particle bases the proton and neutron Fermi surfaces are well contained in the
model space. The Bonn-A G-matrix has been used as the starting point for the 
nucleon-nucleon interaction and it has been renormalized in the standard way 
\cite{Suhonen1988b, Suhonen1993a}: The quasi-particles are treated in the BCS 
formalism and the pairing matrix elements are scaled by a common strength parameter, 
separately for protons and neutrons. In practice these factors are fitted such that 
the lowest quasi-particle energies obtained from the BCS match the experimentally 
deduced pairing gaps for protons and neutrons respectively. For closed major 
shells the pairing strength parameters were taken from the closest even-even neighbor.

The wave functions of the $1^+$ states of the intermediate nuclei have been 
produced by using the pnQRPA with the particle-hole and particle-particle degrees of 
freedom \cite{Vogel1986} included. The particle-hole and particle-particle parts of 
the proton-neutron two-body interaction are separately scaled by the particle-hole 
($g_\text{ph}$) and particle-particle ($g_\text{pp}$) parameters. The particle-hole 
parameter affects the position of the Gamow-Teller giant resonance (GTGR) in the
odd-odd nucleus and its 
value is fixed by the available systematics \cite{Suhonen2007} on the location of 
the resonance:
\begin{equation}
\label{eq:gtgr}
\Delta E_{\rm GT} = E(1^+_{\rm GTGR}) - E(0^+_{\rm gs}) = \Big[ 1.444\left(Z+1/2\right)
A^{-1/3} - 30.0\big( N-Z-2\big)A^{-1} + 5.57\Big]\,\textrm{MeV} .
\end{equation}
The difference $\Delta E_{\rm GT}$ between the GTGR and the ground state of the 
neighboring even-even reference nucleus thus depends on the proton and neutron numbers 
$(Z,N)$ of the reference nucleus, as well as on its mass number. 
In practice, both the measured and the computed GTGR have a width and their location are determined by the centroid (weighted average) of the strengths associated with the individual $1^+$ states comprising the GTGR. In a pnQRPA calculation the difference $E(1^+_\text{GTGR})-E(0^+_\text{gs})$ in Eq.~\eqref{eq:gtgr} gives the empirical location of the centroid of the GTGR which has to be matched by the centroid of the pnQRPA computed strengths of the $1^+$ states presumed to belong to the GTGR. The computed centroid depends strongly on the value of the $g_\text{ph}$ parameter and weakly on the choice of the set of $1^+$ states included in the GTGR, the latter introducing an inherent source of error. Throughout our calculations we assume that the value of $g_\text{ph}$ in a given system is determined with a relative error of 15 \% as a source of theoretical uncertainty. This 15 \% error represents a maximum deviation in $g_\text{ph}$ such that the computed centroid of the Gamow-Teller giant resonance is not meaninglessly far from its empirical position as given by Eq.~\eqref{eq:gtgr}.
Throughout we denote with $\gamma_\text{ph}$ the normalized value of the particle-hole parameter with respect to the value determined through the GTGR. The determination of the values of $g_\text{pp}$, together with the axial-vector coupling constant $g_\text{A}$ is presented below.

\begin{figure}[t!]
\centering
\subfloat[]{
	\includegraphics[width=0.49\columnwidth]{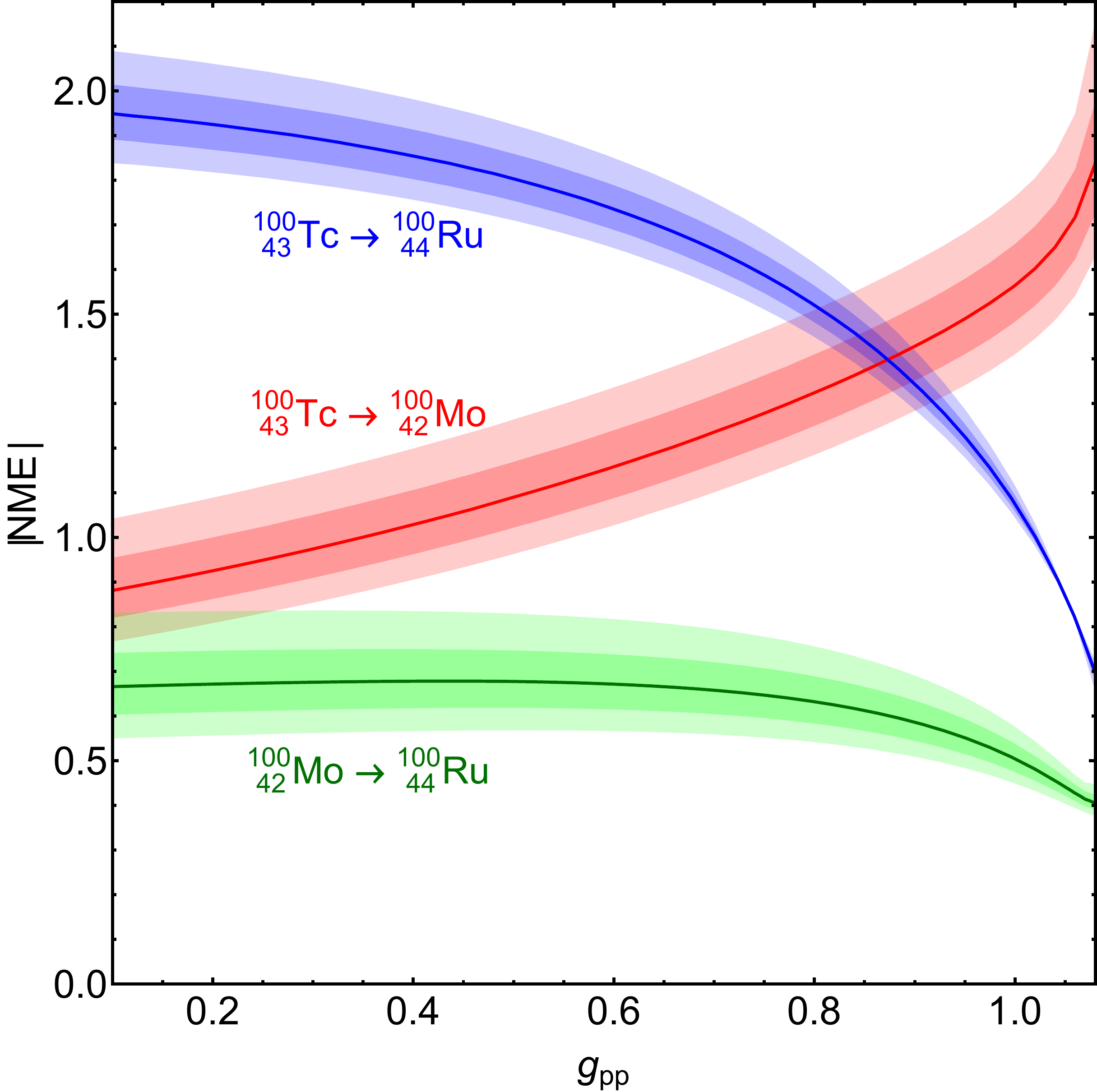}
	\label{fig:nme_gpp_a}
}
\subfloat[]{
	\includegraphics[width=0.49\columnwidth]{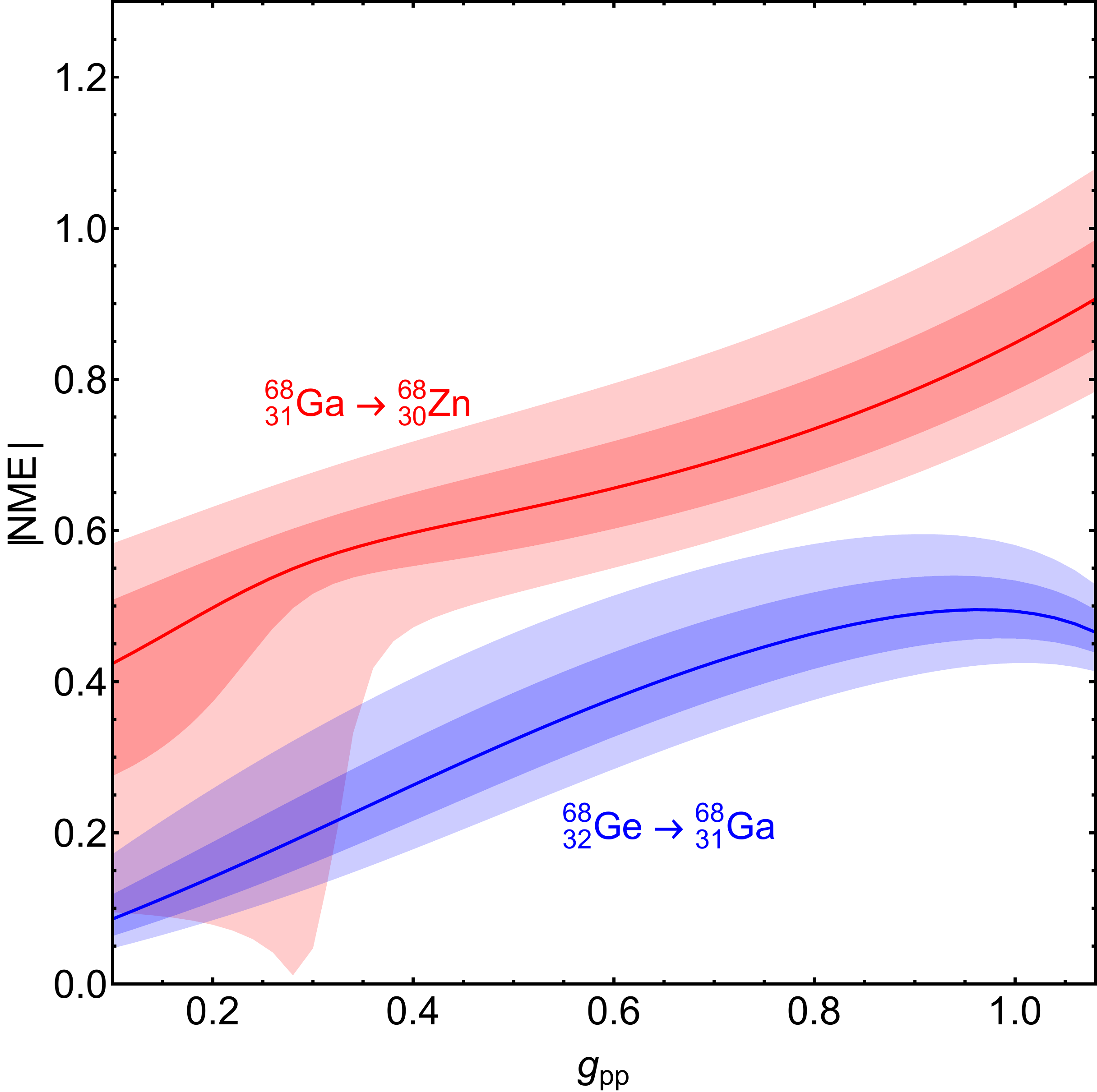}
	\label{fig:nme_gpp_b}
}
\caption{(a) Nuclear matrix elements for the single beta decays 
$^{100}_{\ 43}\text{Tc} \to\,^{100}_{\ 42}\text{Mo}$ (red) and 
$^{100}_{\ 43}\text{Tc} \to\,^{102}_{\ 44}\text{Ru}$ (blue) and the 
$2\nu\beta\beta$ decay $^{102}_{\ 42}\text{Mo} \to\,^{102}_{\ 44}\text{Ru}$ (green) as a 
function of $g_\text{pp}$. The colored bands show the uncertainty of the matrix 
elements with a 15\% variation and a 30\% variation of $g_\text{ph}$ around its value derived from the GTGR.\\(b) As before but for 
the single beta decays $^{68}_{31}\text{Ga} \to\,^{68}_{30}\text{Zn}$ (red) and 
$^{68}_{32}\text{Ge} \to\,^{68}_{31}\text{Ga}$ (blue).}
\label{fig:nme_gpp}
\end{figure}
As an example, Fig.~\ref{fig:schemeMo} schematically shows the energy levels and decay 
characteristics of a triplet of isobars $^{100}_{\ 42}\text{Mo}$, $^{100}_{\ 43}\text{Tc}$ 
and $^{100}_{\ 44}\text{Ru}$. Fig.~\ref{fig:nme_gpp_a} displays the nuclear 
matrix elements of the single beta decays 
$^{100}_{\ 43}\text{Tc} \to\,^{100}_{\ 42}\text{Mo}$ and 
$^{100}_{\ 43}\text{Tc} \to\,^{102}_{\ 44}\text{Ru}$ and the $2\nu\beta\beta$ decay 
$^{102}_{\ 42}\text{Mo} \to\,^{102}_{\ 44}\text{Ru}$ as functions of $g_\text{pp}$. 
The $1\sigma$ and $2\sigma$ uncertainties due to a variation of the parameter 
$g_\text{ph}$ around its value determined by the GTGR are shown using the colored bands. 
The dependence of the single beta NMEs shows a typical behavior seen in many 
triplets where one NME increases whereas the other decreases. This has the effect 
that the dependence of the product of the NMEs on $g_\text{pp}$ can become rather weak 
and consequently the value of $g_\text{A}$ can be extracted from the product of the 
$\log ft$ values separately. This behavior is not universal, as exemplified in 
Fig.~\ref{fig:nme_gpp_b} showing the NMEs of the processes 
$^{68}_{31}\text{Ga} \to\,^{68}_{30}\text{Zn}$ and 
$^{68}_{32}\text{Ge} \to\,^{68}_{31}\text{Ga}$ ($2\nu\beta\beta$ decay is not possible 
here). Here, both NMEs rise with $g_\text{pp}$ which will couple the determination 
of $g_\text{A}$ and $g_\text{pp}$ as is discussed in Sec.~\ref{sec:quenching}. 
As can be seen, a variation of $g_\text{ph}$ at the 15\% level generically has an 
effect on the matrix elements of the same order, depending on the isotopes involved.

\begin{figure}[t!]
\centering
\includegraphics[width=0.55\columnwidth]{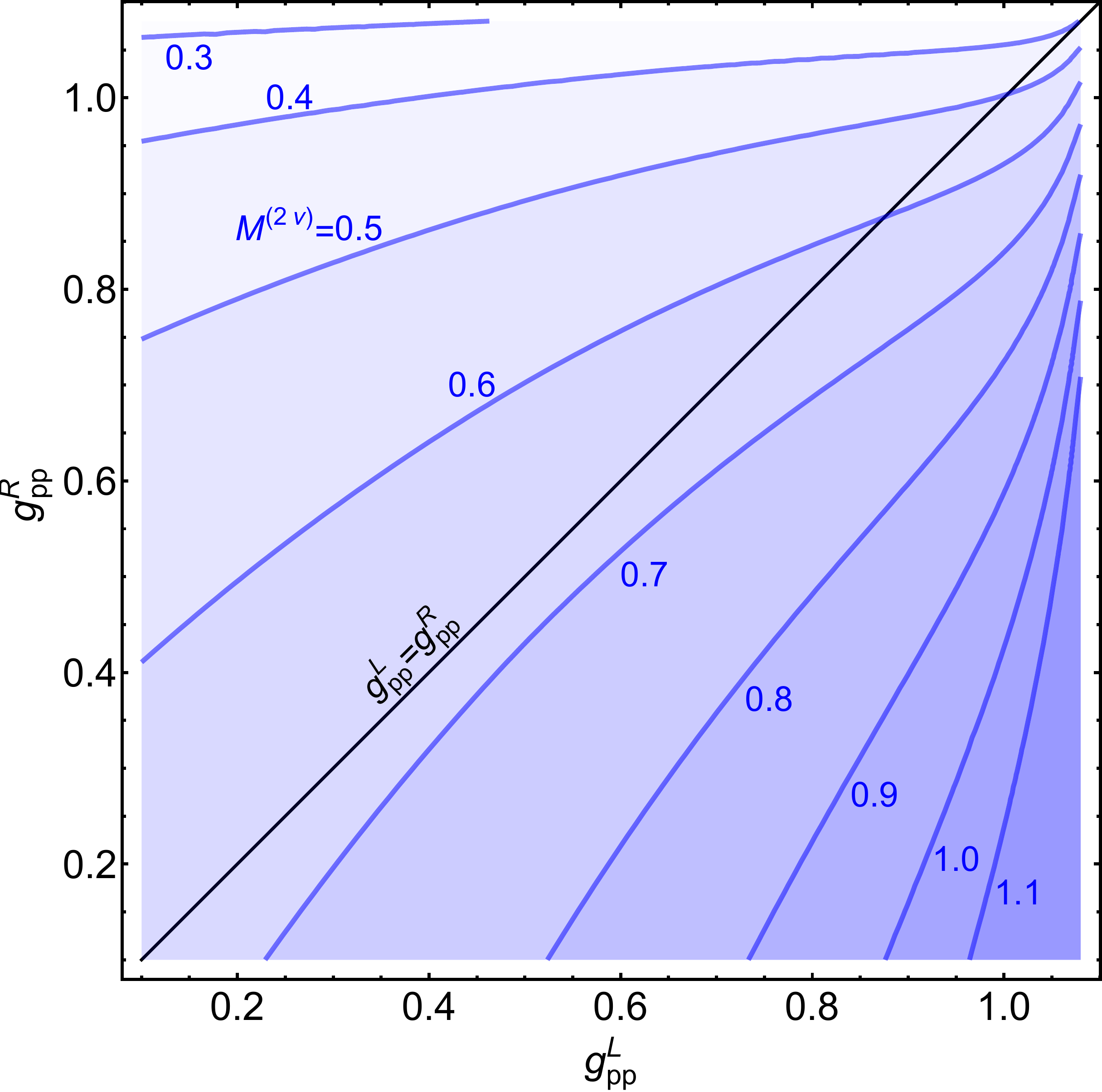}
\caption{Nuclear matrix element for the $2\nu\beta\beta$ decay 
$^{100}_{\ 42}\text{Mo} \to\,^{100}_{\ 44}\text{Ru}$ as a function of the left-leg and 
right-leg particle-particle parameters $g_\text{pp}^L$ and $g_\text{pp}^R$, respectively. 
The iso-curves indicate constant NME values as shown. The particle-hole parameters 
are set at their GTGR values, $\gamma_\text{ph}^L = \gamma_\text{ph}^R = 1$.}
\label{fig:nme_gpp_gpp}
\end{figure}
In Fig.~\ref{fig:nme_gpp} both the left-leg and the right-leg NMEs were treated as 
depending on the same $g_\text{pp}$ as is assumed in most analyses. As discussed 
above, we in turn treat the transitions independently, each depending on separate 
parameters, $g_\text{pp}^L$ and $g_\text{pp}^R$. This has the immediate effect that the 
beta decay/EC processes become statistically independent because they now depend on 
different parameters. In addition, the NME of the $2\nu\beta\beta$ decay (if allowed 
within a given triplet) now becomes a function of both $g_\text{pp}^L$ and 
$g_\text{pp}^R$. Fig.~\ref{fig:nme_gpp_gpp} illustrates the dependence of the NME 
for the $2\nu\beta\beta$ decay $^{102}_{\ 42}\text{Mo} \to\,^{102}_{\ 44}\text{Ru}$. 
The iso-curves indicate constant values for the NME as shown. The plot demonstrates 
that the left-leg and right-leg NMEs are correlated such that the $2\nu\beta\beta$ 
remains approximately constant if both depend on the same $g_\text{pp} \lesssim 0.7$ 
(along the diagonal, also compare with Fig.~\ref{fig:nme_gpp_a}). This degeneracy 
is lifted if $g_\text{pp}^L$ and $g_\text{pp}^R$ are allowed to vary independently. 
The dependence on the particle-hole parameters, which we will in turn also treat as 
independent values, is neglected, and they are set at their GTGR values, 
$\gamma_\text{ph}^L = \gamma_\text{ph}^R = 1$.

\setlength{\tabcolsep}{4.5pt}
\renewcommand{\arraystretch}{1.4}
\begin{table}[t!]
\centering
\begin{tabular}{rr|rcrcr|ll|ll}
\hline
$A$ & $Z_0$ & \multicolumn{5}{|c|}{Triplet} & \multicolumn{1}{c}{$\log ft^\text{exp}_L$} & 
\multicolumn{1}{c|}{$\log ft^\text{exp}_R$} & \multicolumn{1}{c}{$g_\text{A}^\text{fit}$} & 
\multicolumn{1}{c}{$g_\text{pp}^\text{fit}$} \\
\hline
62  & 28 & Ni$(0^+)$ & $\gets$ & Cu$(1^+)$ & $\gets$ & Zn$(0^+)$ & 
$5.1521\pm 0.0014$ & $5.0117\pm 0.0010$ & $	0.75	^{+	0.21	}_{-	0.01	}$	
& $	1.10	^{+	0.01	}_{-	0.40	}$ \\
64  & 28 & Ni$(0^+)$ & $\gets$ & Cu$(1^+)$ & $\to$ & Zn$(0^+)$ & 
$4.9931\pm 0.0022$ & $5.3095\pm 0.0038$ 
& $\mathit{0.81	^{+	0.11	}_{-	0.04	}}$	
& $\mathit{0.92	^{+	0.09	}_{-	0.19	}}$\\
66  & 28 & Ni$(0^+)$ & $\to$ & Cu$(1^+)$ & $\to$ & Zn$(0^+)$ & 
$4.2754\pm 0.0094$ & $5.3394\pm 0.0013$ 
& $	0.93	^{+	0.24	}_{-	0.01	}$	
& $	0.82	^{+	0.04	}_{-	0.34	}$\\
68  & 29 & Cu$(1^+)$ & $\to$ & Zn$(0^+)$ & $\gets$ & Ga$(1^+)$ & 
$5.7716\pm 0.0085$ & $5.1918\pm 0.0012$ 
& $	0.70	^{+	0.06	}_{-	0.09	}$	
& $	0.19	^{+	0.02	}_{-	0.01	}$\\
68  & 30 & Zn$(0^+)$ & $\gets$ & Ga$(1^+)$ & $\gets$ & Ge$(0^+)$  
& $5.1918\pm 0.0012$ & $4.9955\pm 0.0224$ 
& $\mathit{0.50	^{+	0.08	}_{-	0.04	}}$	
& $\mathit{0.78	^{+	0.13	}_{-	0.21	}}$\\
70  & 29 & Cu$(1^+)$ & $\to$ & \underline{Zn}$(0^+)$ & $\gets$ & Ga$(1^+)$  &
$5.4317\pm 0.0138$ & $4.7443\pm 0.0640$ 
& {\textbf -}	& {\textbf -} \\
70  & 30 & \underline{Zn}$(0^+)$ & $\gets$ & Ga$(1^+)$ & $\to$ & Ge$(0^+)$ & 
$4.7443\pm 0.0640$ & $5.1021\pm 0.0018$ 
& $	1.10	^{+	0.13	}_{-	0.26	}$	
& $	0.43	^{+	0.29	}_{-	0.15	}$\\
78  & 34 & Se$(0^+)$ & $\gets$ &Br$(1^+)$ & $\to$ & Kr$(0^+)$ &
$4.7460\pm 0.0040$ & $>5.50\pm 0.01$ 
& $	0.42	^{+	0.02	}_{-	0.04	}$	
& $	1.00	^{+	0.04	}_{-	0.02	}$\\
80  & 33 & As$(1^+)$ & $\to$ & \underline{Se}$(0^+)$ & $\gets$ & Br$(1^+)$ & 
$5.7460\pm 0.0099$ & $4.6868\pm 0.0123$ 
& $	0.98	^{+	0.21	}_{-	0.08	}$	& $	0.34	^{+	0.04	}_{-	0.11	}$\\
80  & 34 & \underline{Se}$(0^+)$ & $\gets$ & Br$(1^+)$ & $\to$ & Kr$(0^+)$ & 
$4.6868\pm 0.0123$ & $5.4953\pm 0.0024$ 
& $\mathit{0.90	^{+	0.33	}_{-	0.07	}}$	
& $\mathit{0.48	^{+	0.06	}_{-	0.23	}}$\\
80  & 35 & Br$(1^+)$ & $\to$ & Kr$(0^+)$ & $\gets$ & Rb$(1^+)$ & 
$5.4953\pm 0.0024$ & $4.9208\pm 0.0514$ 
& $\mathbf{1.40}$	& $\mathbf{0.27}$ \\
98  & 39 & Y$(1^+)$ & $\to$ & Zr$(0^+)$ & $\to$ & Nb$(1^+)$ & 
$5.3740\pm 0.1660$ & $4.1762\pm 0.0170$ 
& $	0.53	^{+	0.05	}_{-	0.03	}$	
& $	0.68	^{+	0.06	}_{-	0.11	}$\\
100 & 41 & Nb$(1^+)$ & $\to$ & \underline{Mo}$(0^+)$ & $\gets$ & Tc$(1^+)$ & 
$5.1622\pm 0.0586$ & $4.4047\pm 0.2414$ 
& $	0.61	^{+	0.14	}_{-	0.15	}$	& $	0.89	^{+	0.06	}_{-	0.08	}$\\
100 & 42 & \underline{Mo}$(0^+)$ & $\gets$ & Tc$(1^+)$ & $\to$ & Ru$(0^+)$ & 
$4.4047\pm 0.2414$ & $4.6063\pm 0.0054$ 
& $	0.56	^{+	0.09	}_{-	0.12	}$	
& $	0.96	^{+	0.05	}_{-	0.16	}$\\
102 & 42 & Mo$(0^+)$ & $\to$ & Tc$(1^+)$ & $\to$ & Ru$(0^+)$ & 
$4.2079\pm 0.0362$ & $4.8001\pm 0.0129$ 
& $	0.41	^{+	0.02	}_{-	0.02	}$	& $	0.68	^{+	0.05	}_{-	0.06	}$\\
104 & 44 & \underline{Ru}$(0^+)$ & $\gets$ & Rh$(1^+)$ & $\to$ & Pd$(0^+)$ & 
$4.3246\pm 0.1030$ & $4.5555\pm 0.0056$ 
& $	0.59	^{+	0.05	}_{-	0.07	}$	& $	0.92	^{+	0.03	}_{-	0.06	}$\\
106 & 45 & Rh$(1^+)$ & $\to$ & Pd$(0^+)$ & $\gets$ & Ag$(1^+)$ & 
$5.1899\pm 0.0060$ & $4.9148\pm 0.0035$ 
& $	0.40	^{+	0.02	}_{-	0.02	}$	& $	0.87	^{+	0.01	}_{-	0.01	}$\\
106 & 46 & Pd$(0^+)$ & $\gets$ & Ag$(1^+)$ & $\to$ & Cd$(0^+)$ & 
$4.9148\pm 0.0035$ & $>4.18\pm 0.25$ 
& $	0.36	^{+	0.25	}_{-	0.04	}$	& $	1.00	^{+	0.13	}_{-	0.72	}$\\
108 & 44 & Ru$(0^+)$ & $\to$ & Rh$(1^+)$ & $\to$ & Pd$(0^+)$ & 
$4.4885\pm 0.0223$ & $5.5440\pm 0.0480$ 
& $	0.27	^{+	0.01	}_{-	0.02	}$	& $	0.69	^{+	0.04	}_{-	0.06	}$\\
108 & 45 & Rh$(1^+)$ & $\to$ & Pd$(0^+)$ & $\gets$ & Ag$(1^+)$ & 
$5.5440\pm 0.0480$ & $4.7085\pm 0.0372$ 
& $	0.43	^{+	0.03	}_{-	0.05	}$	& $	0.86	^{+	0.19	}_{-	0.01	}$\\
108 & 46 & Pd$(0^+)$ & $\gets$ & Ag$(1^+)$ & $\to$ & Cd$(0^+)$ & 
$4.7085\pm 0.0372$ & $4.4410\pm 0.0080$ 
& $	0.49	^{+	0.02	}_{-	0.02	}$	& $	0.67	^{+	0.05	}_{-	0.08	}$\\
110 & 46 & \underline{Pd}$(0^+)$ & $\gets$ & Ag$(1^+)$ & $\to$ & Cd$(0^+)$ & 
$4.0963\pm 0.0887$ & $4.6762\pm 0.0021$ 
& $	0.77	^{+	0.06	}_{-	0.08	}$	& $	0.87	^{+	0.02	}_{-	0.04	}$\\
112 & 48 & Cd$(0^+)$ & $\gets$ & In$(1^+)$ & $\to$ & Sn$(0^+)$ & 
$4.6342\pm 0.0378$ & $4.1515\pm 0.0497$ 
& $	0.70	^{+	0.04	}_{-	0.03	}$	& $	0.61	^{+	0.07	}_{-	0.11	}$\\
114 & 46 &  Pd$(0^+)$ & $\to$ & Ag$(1^+)$ & $\to$ & Cd$(0^+)$ &
$4.2124\pm 0.0153$ & $5.1008\pm 0.0096$ 
& $	0.51	^{+	0.03	}_{-	0.03	}$	& $	0.49	^{+	0.06	}_{-	0.09	}$\\
114 & 47 & Ag$(1^+)$ & $\to$ & \underline{Cd}$(0^+)$ & $\gets$ & In$(1^+)$ & 
$5.1008\pm 0.0096$ & $4.8877\pm 0.1470$ 
& $	0.54	^{+	0.06	}_{-	0.07	}$	& $	0.54	^{+	0.08	}_{-	0.13	}$\\
114 & 48 & \underline{Cd}$(0^+)$ & $\gets$ & In$(1^+)$ & $\to$ & Sn$(0^+)$ & 
$4.8877\pm 0.1470$ & $4.4856\pm 0.0010$ 
& $\mathit{0.61	^{+	0.06	}_{-	0.01	}}$	
& $\mathit{0.46	^{+	0.15	}_{-	0.01	}}$\\
\hline
\end{tabular}
\caption{Characteristics of the $\beta^+$/EC and $\beta^-$ decays in isobaric triplets 
within the mass range $A = 62 - 114$ studied in the present paper. An isobaric triplet 
is identified by the mass number $A$ and the lowest atomic number $Z_0$ among the 
three isotopes. The isotopes in the triplets are indicated along with their spin $J$ 
and parity $\pi$, $(J^\pi)$. The arrows denote the direction of the relevant 
$\beta^+$/EC, $\beta^-$ decay. The experimentally determined comparative half-lives 
of the left and right transition are given as $\log ft_L$ and $\log ft_R$, respectively. 
$2\nu\beta\beta$ decaying isotopes are underlined. The values of $g_\text{A}$ and 
$g_\text{pp}$ are determined in the triplet fit described in 
Sec.~\ref{sec:results_triplets}. Cases in which the best fit $\chi^2_\text{min}$ is in 
the range $[0.5, 2.2]$, indicating slight incompatibility with data, are highlighted with italic numbers. Cases with stronger discrepancy are highlighted in bold. In all 
other cases a $\chi^2_\text{min} = 0$ (within numerical tolerance) was found.}
\label{tab:triplets} 
\end{table}
\begin{table}[t!]
\centering
\begin{tabular}{rr|rcrcr|ll|ll}
\hline
$A$ & $Z_0$ & \multicolumn{5}{|c|}{Triplet} & \multicolumn{1}{c}{$\log ft^\text{exp}_L$} 
& \multicolumn{1}{c|}{$\log ft^\text{exp}_R$} & \multicolumn{1}{c}{$g_\text{A}^\text{fit}$} 
& \multicolumn{1}{c}{$g_\text{pp}^\text{fit}$} \\
\hline
116 & 48 & \underline{Cd}$(0^+)$ & $\gets$ & In$(1^+)$ & $\to$ & Sn$(0^+)$ & 
$4.4508\pm 0.1160$ & $4.6839\pm 0.0025$ 
& $	0.84	^{+	0.08	}_{-	0.08	}$	& $	0.65	^{+	0.07	}_{-	0.11	}$\\
118 & 48 & Cd$(0^+)$ & $\to$ & In$(1^+)$ & $\to$ & Sn$(0^+)$ & 
$3.9218\pm 0.0629$ & $4.8147\pm 0.0263$ 
& $	0.88	^{+	0.09	}_{-	0.07	}$	& $	0.75	^{+	0.04	}_{-	0.09	}$\\
118 & 49 & In$(1^+)$ & $\to$ & Sn$(0^+)$ & $\gets$ & Sb$(1^+)$ & 
$4.8147\pm 0.0263$ & $4.5152\pm 0.0122$ 
& $	0.77	^{+	0.05	}_{-	0.06	}$	& $	0.65	^{+	0.03	}_{-	0.04	}$\\
118 & 50 & Sn$(0^+)$ & $\gets$ & Sb$(1^+)$ & $\gets$ & Te$(0^+)$ & 
$4.5152\pm 0.0122$ & $4.9749\pm 0.0579$ 
& $	0.77	^{+	0.06	}_{-	0.05	}$	& $	0.65	^{+	0.04	}_{-	0.14	}$\\
120 & 48 & Cd$(0^+)$ & $\to$ & In$(1^+)$ & $\to$ & Sn$(0^+)$ & 
$4.0996\pm 0.0433$ & $5.0483\pm 0.0183$ 
& $	0.74	^{+	0.07	}_{-	0.05	}$	& $	0.77	^{+	0.04	}_{-	0.08	}$\\
120 & 49 & In$(1^+)$ & $\to$ & Sn$(0^+)$ & $\gets$ & Sb$(1^+)$ & 
$5.0483\pm 0.0183$ & $4.5220\pm 0.0048$ 
& $	0.71	^{+	0.06	}_{-	0.06	}$	& $	0.74	^{+	0.03	}_{-	0.03	}$\\
122 & 48 & Cd$(0^+)$ & $\to$ & In$(1^+)$ & $\to$ & \underline{Sn}$(0^+)$ & 
$3.9717\pm 0.0451$ & $5.1362\pm 0.0894$ 
& $	0.82	^{+	0.09	}_{-	0.06	}$	& $	0.85	^{+	0.05	}_{-	0.08	}$\\
122 & 52 & Te$(0^+)$ & $\gets$ & I$(1^+)$ & $\gets$ & Xe$(0^+)$ & 
$4.9323\pm 0.0077$ & $5.1804\pm 0.0154$ 
& $	0.50	^{+	0.04	}_{-	0.03	}$	& $	0.60	^{+	0.05	}_{-	0.12	}$\\
122 & 53 & I$(1^+)$ & $\gets$ & Xe$(0^+)$ & $\gets$ & Cs$(1^+)$ & 
$5.1804\pm 0.0154$ & $5.3606\pm 0.0102$ 
& $	0.43	^{+	0.04	}_{-	0.03	}$	& $	0.36	^{+	0.02	}_{-	0.02	}$\\
124 & 54 & Xe$(0^+)$ & $\gets$ & Cs$(1^+)$ & $\gets$ & Ba$(0^+)$  & 
$5.0750\pm 0.0080$ & $5.2074\pm 0.0216$ 
& $	0.39	^{+	0.03	}_{-	0.02	}$	& $	0.71	^{+	0.03	}_{-	0.06	}$\\
126 & 54 & Xe$(0^+)$ & $\gets$ & Cs$(1^+)$ & $\gets$ & Ba$(0^+)$  & 
$5.0492\pm 0.0084$ & $5.3577\pm 0.0135$ 
& $	0.44	^{+	0.03	}_{-	0.03	}$	& $	0.67	^{+	0.04	}_{-	0.08	}$\\
128 & 52 & \underline{Te}$(0^+)$ & $\gets$ & I$(1^+)$ & $\to$ & Xe$(0^+)$  & 
$5.0439\pm 0.0514$ & $6.0825\pm 0.0055$ 
& $	0.55	^{+	0.08	}_{-	0.03	}$	& $	0.10	^{+	0.26	}_{-	0.05	}$\\
128 & 53 & I$(1^+)$ & $\to$ & Xe$(0^+)$ & $\gets$ & Cs$(1^+)$  & 
$6.0825\pm 0.0055$ & $4.8255\pm 0.0036$ 
& $	0.68	^{+	0.09	}_{-	0.07	}$	& $	0.43	^{+	0.01	}_{-	0.01	}$\\
128 & 54 & Xe$(0^+)$ & $\gets$ & Cs$(1^+)$ & $\gets$ & Ba$(0^+)$ & 
$4.8255\pm 0.0036$ & $5.3973\pm 0.0235$ 
& $	0.58	^{+	0.05	}_{-	0.05	}$	& $	0.65	^{+	0.04	}_{-	0.09	}$\\
130 & 54 & Xe$(0^+)$ & $\gets$ & Cs$(1^+)$ & $\to$ & Ba$(0^+)$ & 
$5.0654\pm 0.0049$ & $5.1314\pm 0.0692$ 
& $\textbf{0.78}$	& $\textbf{0.10}$\\
134 & 56 & Ba$(0^+)$ & $\gets$ & La$(1^+)$ & $\gets$ & Ce$(0^+)$ & 
$4.8703\pm 0.0154$ & $5.1920\pm 0.0790$ 
& $	0.73	^{+	0.07	}_{-	0.06	}$	& $	0.34	^{+	0.12	}_{-	0.13	}$\\
138 & 58 & Ce$(0^+)$ & $\gets$ & Pr$(1^+)$ & $\gets$ & Nd$(0^+)$ & 
$4.5880\pm 0.0160$ & $5.0934\pm 0.0422$ 
& $	0.98	^{+	0.08	}_{-	0.08	}$	& $	0.47	^{+	0.07	}_{-	0.14	}$\\
140 & 58 & Ce$(0^+)$ & $\gets$ & Pr$(1^+)$ & $\gets$ & Nd$(0^+)$ & 
$4.4064\pm 0.0035$ & $5.4279\pm 0.0643$ 
& $	1.00	^{+	0.09	}_{-	0.07	}$	& $	0.46	^{+	0.07	}_{-	0.16	}$\\
140 & 59 & Pr$(1^+)$ & $\gets$ & Nd$(0^+)$ & $\gets$ & Pm$(1^+)$ & 
$5.4279\pm 0.0643$ & $4.3085\pm 0.0129$ 
& $	1.30	^{+	0.06	}_{-	0.15	}$	& $	0.61	^{+	0.02	}_{-	0.03	}$\\
140 & 60 & Nd$(0^+)$ & $\gets$ & Pm$(1^+)$ & $\gets$ & Sm$(0^+)$ & 
$4.3085\pm 0.0129$ & $4.8933\pm 0.0214$ 
& $	1.20	^{+	0.08	}_{-	0.09	}$	& $	0.66	^{+	0.04	}_{-	0.08	}$\\
140 & 61 & Pm$(1^+)$ & $\gets$ & Sm$(0^+)$ & $\gets$ & Eu$(1^+)$ & 
$4.8933\pm 0.0214$ & $4.3916\pm 0.0142$ 
& $	1.20	^{+	0.10	}_{-	0.13	}$	& $	0.67	^{+	0.01	}_{-	0.01	}$\\
140 & 62 & Sm$(0^+)$ & $\gets$ & Eu$(1^+)$ & $\gets$ & Gd$(0^+)$ & 
$4.3916\pm 0.0142$ & $4.5357\pm 0.0266$ 
& $	1.10	^{+	0.08	}_{-	0.07	}$	& $	0.74	^{+	0.03	}_{-	0.07	}$\\
142 & 60 & Nd$(0^+)$ & $\gets$ & Pm$(1^+)$ & $\gets$ & Sm$(0^+)$ & 
$4.4687\pm 0.0183$ & $5.1656\pm 0.0151$ 
& $	1.00	^{+	0.08	}_{-	0.06	}$	& $	0.45	^{+	0.08	}_{-	0.12	}$\\
142 & 61 & Pm$(1^+)$ & $\gets$ & Sm$(0^+)$ & $\gets$ & Eu$(1^+)$ & 
$5.1656\pm 0.0151$ & $4.2736\pm 0.0239$ 
& $	1.30	^{+	0.03	}_{-	0.17	}$	& $	0.67	^{+	0.01	}_{-	0.02	}$\\
\hline
\end{tabular}
\caption{As Table~\ref{tab:triplets}, for isobaric triplets in the mass range $A = 116 - 142$.}
\label{tab:triplets2} 
\end{table}
\begin{table}[t!]
\centering
\begin{tabular}{rr|c|cc|cc}
\hline
$A$ & $Z$ & Isotope & $G_{2\nu}$ [yr$^{-1}$]& $[t_{1/2}^{(2\nu)}]_\text{exp}$ [yr] 
& $[t_{1/2}^{(2\nu)}]_\text{triplet}$ [yr] & $[t_{1/2}^{(2\nu)}]_\text{multiplet}$ [yr] \\
\hline
 70	&	30 & Zn	& $1.24\times 10^{-22}$ & -	                             
& $(7.0\pm 4.1)\times 10^{22}$ & {\textbf -} \\
 80	&	34 & Se	& $7.06\times 10^{-29}$ & -	                             &  
 $(2.6\pm 1.7)\times 10^{29}$ & {\textbf -} \\
100	&	42 & Mo	& $3.87\times 10^{-18}$ & $(0.71\pm 0.04)\times 10^{19}$ 
&  $(1.1\pm 0.6)\times 10^{19}$ & $(1.5\pm 0.6)\times 10^{19}$ \\
104	&	44 & Ru	& $3.80\times 10^{-21}$ & -	                             
& $(7.8\pm 1.7)\times 10^{21}$ & $(4.3\pm 3.7)\times 10^{21}$ \\
110	&	46 & Pd	& $1.64\times 10^{-19}$ & -	                             
& $(1.5\pm 0.3)\times 10^{20}$ & $(1.3\pm 0.4)\times 10^{20}$ \\
114	&	48 & Cd	& $6.09\times 10^{-24}$ & -	                             
& $(7.0\pm 1.2)\times 10^{24}$ & $\mathit{(7.5\pm 1.2)\times 10^{24}}$ \\
116	&	48 & Cd	& $3.27\times 10^{-18}$ & $(2.85\pm 0.15)\times 10^{19}$ 
& $(1.3\pm 0.3)\times 10^{19}$ & $(1.2\pm 0.3)\times 10^{19}$ \\
122	&	50 & Sn	& $4.45\times 10^{-25}$ & -	                             
& $(2.2\pm 2.0)\times 10^{27}$ & $\mathit{(1.7\pm 0.7)\times 10^{27}}$ \\
128	&	52 & Te	& $3.61\times 10^{-22}$ & $(2.00\pm 0.30)\times 10^{24}$ 
& $(0.8\pm 0.2)\times 10^{24}$ & $\mathit{(1.0\pm 0.2)\times 10^{24}}$ \\
\hline
\end{tabular}
\caption{Characteristics of $2\nu\beta\beta$ isotopes studied in the present paper. 
The phase space factors $G_{2\nu}$ were calculated using the formalism of \cite{REPORT}. 
The experimental half-lives were reported in \cite{Barabash:2013aya}. The theoretically 
determined values $[t_{1/2}^{(2\nu)}]_\text{triplet}$ and 
$[t_{1/2}^{(2\nu)}]_\text{multiplet}$ are the \emph{predictions} for the $2\nu\beta\beta$ 
half-lifes based on the triplet and multiplet single beta/EC fits described in 
Secs.~\ref{sec:results_triplets} and \ref{sec:results_isobars}, respectively. 
The italicized half-lifes are derived in multiplet fits with a slight 
tension between data and theory. For $^{70}$Zn and $^{80}$Se, no meaningful multiplet 
result could be derived due to the insufficient quality of the underlying fits.}
\label{tab:2vbbresults} 
\end{table}

\section{Statistical analysis}
\label{sec:quenching}

\subsection{Quenching of the axial-vector coupling constant}

At this stage it is worth pointing to some other earlier works devoted to the 
determination of the effective value of $g_\text{A}$ in calculations using the 
pnQRPA model or other models. A strongly reduced effective value of 
$g_\text{A} \approx 0.6$ was reported in the shell-model calculations 
\cite{Juodagalvis2005} in the mass $A=90-97$ region. In a more recent shell-model 
study \cite{Caurier2012} values of about $g_\text{A} \sim 0.7$ were obtained in the 
mass region $A=128, 130$ and an even a stronger quenching of $g_\text{A} = 0.56$ was obtained for $A = 136$. 
The first analysis performed in the pnQRPA model was done in Ref.~\cite{Faessler2008} where 
both the beta-decay and the $2\nu\beta\beta$ decay data were analyzed for the 
$A = 100, 116$ systems using a least-squares fit to determine the values 
$g_\text{A} = 0.74$ ($A = 100$) and $g_\text{A} = 0.84$ ($A = 116$). It is interesting 
to note that in the first version \cite{Faessler-draft} of Ref.~\cite{Faessler2008} 
also the result $g_\text{A} = 0.39$ for the $A=128$ system was quoted. An approximately 
monotonic behavior of the effective values of $g_\text{A}$ was parametrized in 
Ref.~\cite{Barea2013} by analyzing the magnitudes of NMEs produced by the IBA-2 model. 
Values around $g_\text{A}=0.5$ were obtained. In a later publication \cite{Yoshida2013} 
the interacting boson-fermion-fermion model, IBFFM-2, was adopted and the subsequent 
analyses yielded highly suppressed values of $g_\text{A} \approx 0.3$ for the $A = 128$ nuclei.
Recently a systematic approach to $\beta$ and $2\nu\beta\beta$ decays in the
mass region $A=100-136$ was performed \cite{Pirinen2015}. The suitability to the description
of the global behavior of the $\beta$ and $2\nu\beta\beta$ decays, a linear model and an overall-quenched $g_\text{A} \approx 0.6$ were examined. The present study is
an extension of this work as well as Refs.~\cite{Suhonen2005, Suhonen2013c, Suhonen2014} to a wider mass region and a refinement in the statistical
analysis methods used to extract information on the quenching of $g_\text{A}$ in this
wider region.

The apparently low effective values of $g_\text{A}$ in the pnQRPA  could be attributed to 
missing the contributions of the complex configurations beyond the two-quasiparticle 
(particle-hole) configurations of the pnQRPA (see also Ref.~\cite{Holt2013}). On the other 
hand, it was shown in Ref.~\cite{Kotila2009} for the $2\nu\beta\beta$ decay of $^{76}$Ge and 
in Ref.~\cite{Kotila2010} for the $2\nu\beta\beta$ decay of $^{100}$Mo that the inclusion 
of the four-quasiparticle (two-particle--two-hole) degrees of freedom in a 
higher-QRPA scheme (in this case the pnMAVA (proton-neutron microscopic anharmonic vibrator approach)) does not affect appreciably the low-energy Gamow-Teller properties of pnQRPA. It is yet unclear what is the primary 
reason for the rather low effective values of $g_\text{A}$ and what is the share 
between the model-dependent and model-independent contributions to it. The 
model-independent quenching can be associated with the non-nucleonic, i.e. isobaric 
degrees of freedom in nuclear matter \cite{Bohr1975, Bohr1981}. Contributions to the 
model-dependent quenching come from the limitations in the single-particle models 
space (ISM, IBA-2, IBFFM-2) or the lack of complicated many-nucleon configurations 
(pnQRPA, IBA-2, IBFFM-2). The determination of the effective values of $g_\text{A}$ 
in different theory frameworks is an extremely interesting issue and certainly 
necessitates further investigation in the future.

\subsection{Fitting isobaric triplets}
\label{sec:results_triplets}

The basis of our analyses is provided by the experimental $\log ft$ values of the 
relevant beta decays/EC processes. They are shown in Tabs.~\ref{tab:triplets} and 
\ref{tab:triplets2}, displaying the comparative half-lives $\log ft_L$ and $\log ft_R$ 
for the left-leg decay and the right-leg decay of a given triplet. The comparative half-lives 
were calculated from the experimentally measured half-lives listed in Ref.~\cite{ENSDF} 
incorporating the experimental uncertainty in both the measured decay half-life and 
the $Q$ value. In most cases, the experimental uncertainty is negligible compared to 
the theoretical uncertainties expected to be inherent in nuclear model calculation; 
the errors in the values of the comparative half-lives range between the per mil 
and the 10\% level. In two cases only, a lower limit is known. The range of considered 
isotopes is dictated by the applicability of the theory framework (the quasiparticle 
description for s-d shell nuclei becomes questionable for lighter nuclei) and nature 
(for example, the $1^+$ states are not the ground states in the odd-odd systems in the 
mass gaps $A=$72-76 and $A=$82-96).

As far as available and relevant for our selection of isotopes we also calculate 
the $2\nu\beta\beta$ decay half-lives. The characteristics of $2\nu\beta\beta$ decaying 
isotopes are shown in Table~\ref{tab:2vbbresults} giving the phase space factor 
$G_{2\nu}$, and the experimental half-lives for three of the isotopes. The fitted or 
predicted half-lives are discussed below.

We start by fitting the triplets individually, i.e. we compare the theoretically 
predicted $\log ft$ values of the left-leg and right-leg decays in a triplet of 
Tabs.~\ref{tab:triplets} and \ref{tab:triplets2} with the experimental data. For 
this purpose, we assume that both decays depend on the same pair $g_\text{A}$, 
$g_\text{pp}$. In addition we also include a variation of the $g_\text{ph}$ couplings
independently for each decay, with a $1\sigma$ deviation of 15\% from its GTGR value. 

Because we later work with a larger number of free parameters and in systems that can 
be under-constrained, exactly-constrained or over-constrained, we consistently perform the fitting procedure 
using a straightforward MCMC based on the Metropolis-Hastings 
algorithm \cite{Press:1992zz}. Throughout our calculations, we have verified that the 
uncertainties inherent in the MCMC due to finite sampling etc. are small compared to 
the physical uncertainties. We always use a flat prior in the given fitting 
parameters, i.e. they are randomly selected on a linear scale within a given range. 
We always vary $g_\text{A}$ and $g_\text{pp}$ in the range $[0.1, 1.4]$ and $\gamma_\text{ph}$ between $[0.5, 1.5]$.

\begin{figure}[t!]
\centering
\subfloat[]{\includegraphics[width=0.49\columnwidth]{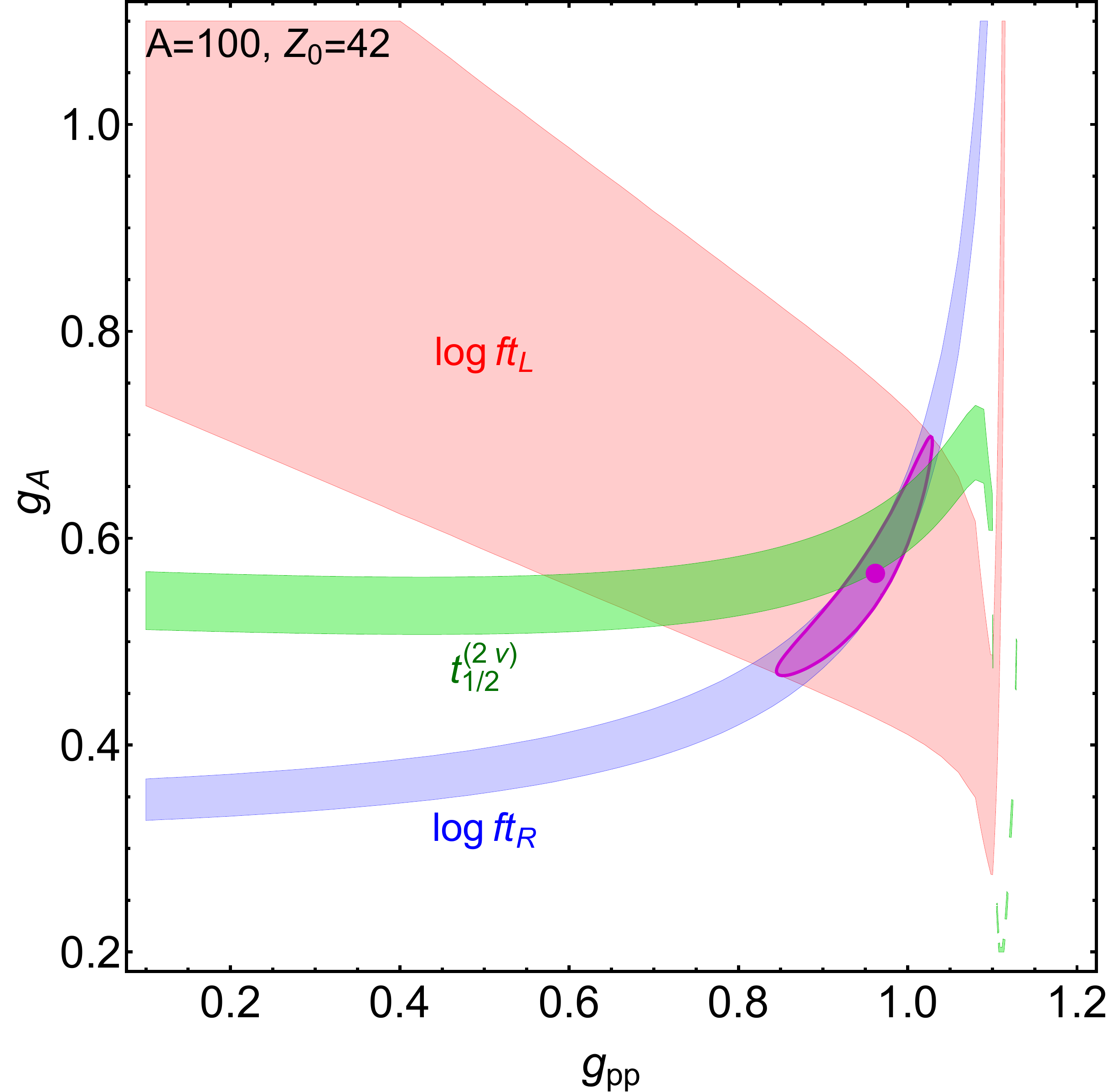}\label{fig:chi2_a}}
\subfloat[]{\includegraphics[width=0.49\columnwidth]{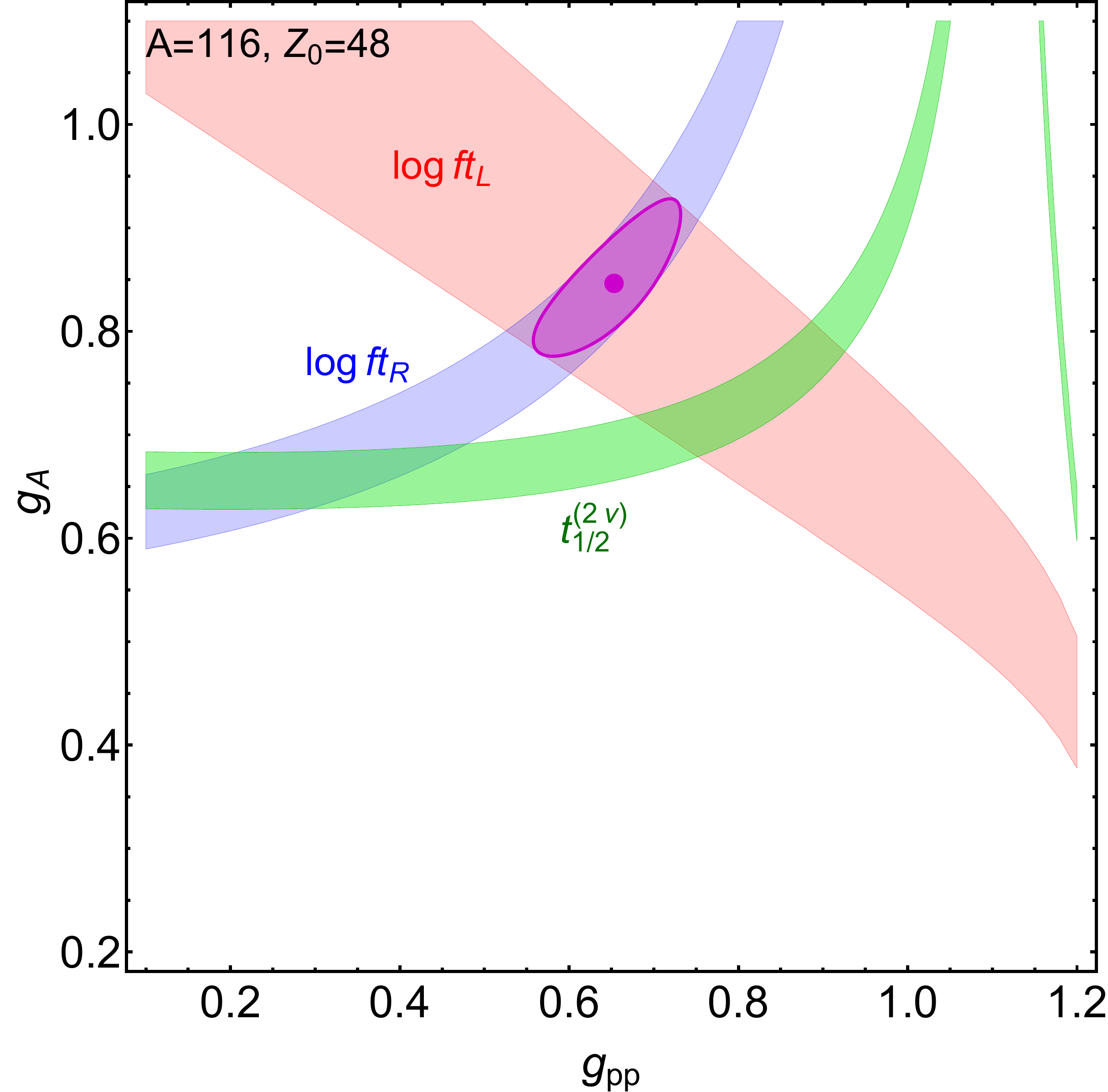}\label{fig:chi2_b}}\\
\subfloat[]{\includegraphics[width=0.49\columnwidth]{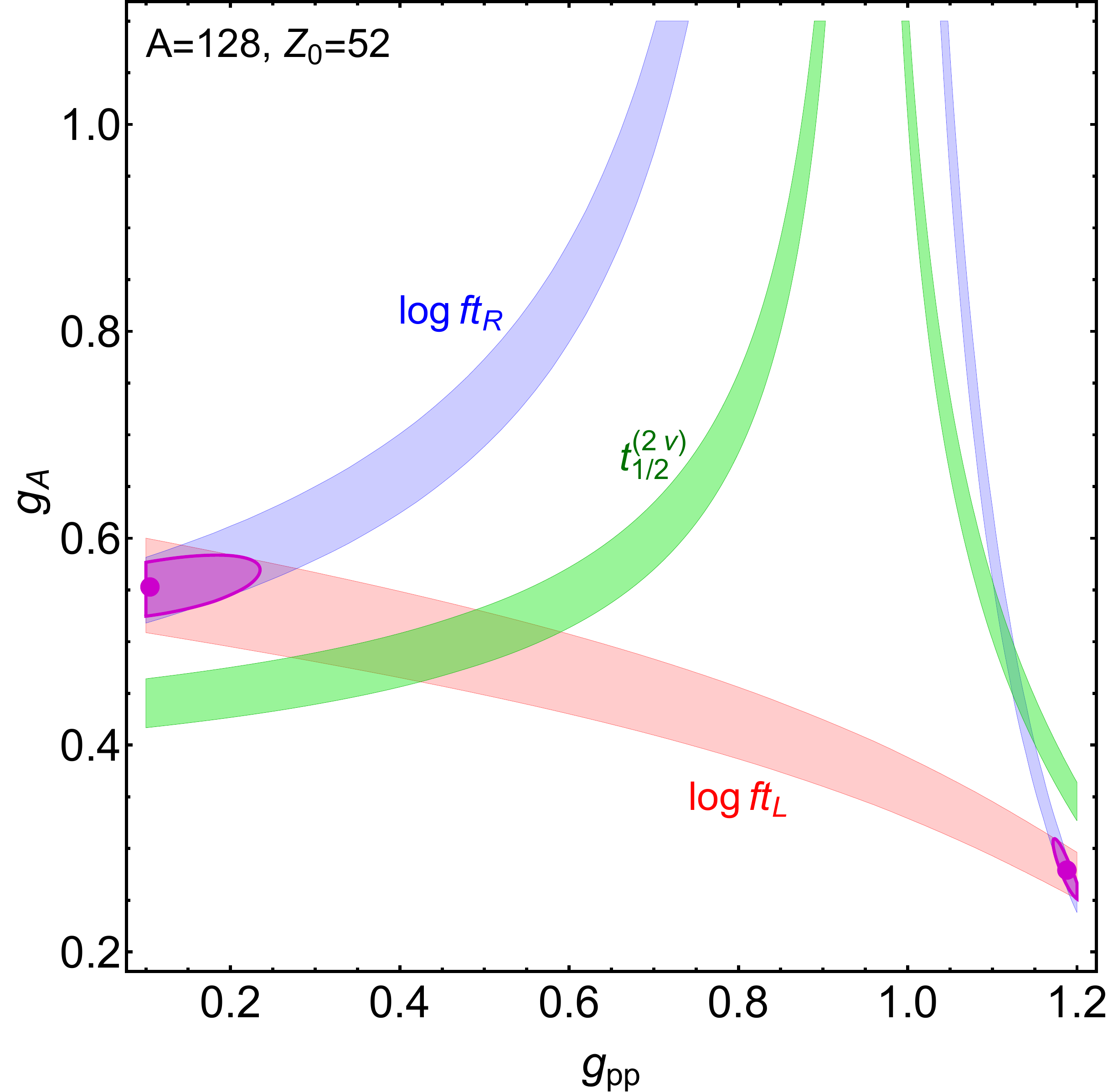}\label{fig:chi2_c}}
\subfloat[]{\includegraphics[width=0.49\columnwidth]{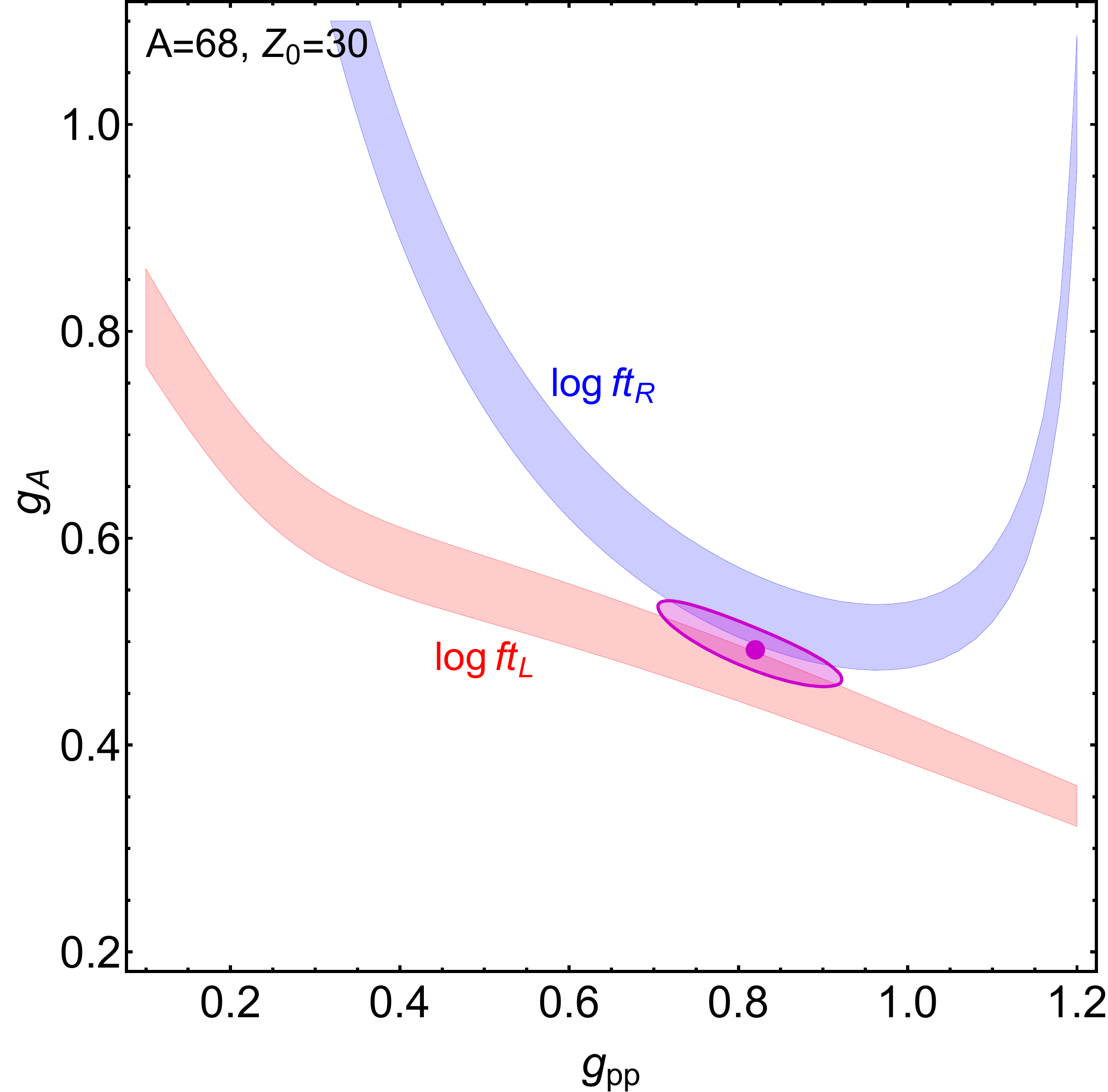}\label{fig:chi2_d}}
\caption{Fitting $g_\text{A}$ and $g_\text{pp}$ in selected isobar triplets: 
$(A,Z_0) = (100,42)$ (a), $(116,48)$ (b), $(128,52)$ (c) and $(68,30)$ (d). The light shaded red and blue bands correspond to the individual $1\sigma$ constraint from the measurement of left-leg and right-leg beta/EC decay, respectively, using the simplified $\chi^2$ in Eq.~\eqref{eq:chi2_triplet_simple}. The dark purple area gives the combined $1\sigma$ parameter area 
with the dot denoting the best-fit. The first three examples contain a measured 
$2\nu\beta\beta$ decaying isotope and the green band gives the corresponding 
parameter space. In addition to the experimental errors from 
Tabs.~\ref{tab:triplets}, \ref{tab:triplets2} and \ref{tab:2vbbresults}, a common 
theoretically induced error of 10\% in the respective observables is included.}
\label{fig:chi2}
\end{figure}
In the current case, the fitting is based on a $\chi^2$ function applied to a given 
triplet of the form
\begin{align}
\label{eq:chi2_triplet}
	\chi^2(g_\text{A}, g_\text{pp}, \gamma_\text{ph}^L, \gamma_\text{ph}^R) &= 
	\left(
	\frac{\log ft^\text{th}_L(g_\text{A}, g_\text{pp}, \gamma_\text{ph}^L) - 
     \log ft^\text{exp}_L}
	{\delta\log ft^\text{exp}_L}
	\right)^2
    \!\!  + \!\!
	\left(
	\frac{\log ft^\text{th}_R(g_\text{A}, g_\text{pp}, \gamma_\text{ph}^R) - 
    \log ft^\text{exp}_R}
	{\delta\log ft^\text{exp}_R}
	\right)^2
	\nonumber\\
	&+
	\left(\frac{\gamma_\text{ph}^L - 1}{\delta\gamma_\text{ph}^L}\right)^2 
	\left\{+ \left(\frac{\gamma_\text{ph}^R - 1}{\delta\gamma_\text{ph}^R}\right)^2 
\right\}.
\end{align}
Here, the experimental $\log ft_{L,R}^\text{exp}$ values along with their experimental 
errors $\delta\log ft_{L,R}^\text{exp}$ are taken from Tabs.~\ref{tab:triplets} and 
\ref{tab:triplets2}. The theoretically determined $\log ft^\text{th}_{L,R}$ are
computed as functions of the fitting parameters $g_\text{A}$ and $g_\text{pp}$. 
In addition, they depend on the variables $\gamma_\text{ph}^{L,R}$ which represent 
the particle-hole parameters relative to values as derived from the energy of the
giant resonance, $\gamma_\text{ph}^{L,R} = g_\text{ph}^{L,R}/[g_\text{ph}^{L,R}]_\text{GTGR}$. 
The last two terms in Eq.~\eqref{eq:chi2_triplet} correspond to using the 
$\gamma_\text{ph}^{L,R}$ as nuisance parameters with best fit values of $1$ and the 
deviations $\delta\gamma_\text{ph}^{L,R} = 0.15$. In the case where only a lower limit 
on the experimental $\log ft$ is known, the corresponding quadratic term in $\chi^2$ is 
replaced by 
$(\text{max}(0, \log ft^\text{th} - \log ft^\text{exp})/\delta\log ft^\text{exp})^2$, 
i.e. a single sided exponential to represent the lower limit. Finally, the MCMC fit is performed 
using the fitness function $P = \exp(-\chi^2/2)$.

With two (non-nuisance) parameters and two constraints, the system is exactly 
determined and a solution with $\chi^2 = 0$ is generically expected. As is shown later, in some cases no consistent solution can be found for the given experimental data. 
As indicated by the curly brackets, for triplets in which the central isotope is 
even-even (i.e. identified by an odd $Z_0$), both decay legs are regulated by the 
same $g_\text{ph}$ and thus the fit is performed with only one nuisance term and 
$\gamma_\text{ph}^L = \gamma_\text{ph}^R$.

Before discussing the results of the numerical fits, we would like to illustrate 
how the experimental data constrains $(g_\text{A}, g_\text{pp})$ in a few examples. 
By omitting the nuisance parameters and shifting their induced uncertainty into a 
theoretical error on the $\log ft$, Eq.~\eqref{eq:chi2_triplet} simplifies to
\begin{align}
\label{eq:chi2_triplet_simple}
	\chi^2(g_\text{A}, g_\text{pp}) &= 
	\frac{(\log ft^\text{th}_L(g_\text{A}, g_\text{pp}) - \log ft^\text{exp}_L)^2}
	{(\delta\log ft^\text{exp}_L)^2 + (\delta\log ft^\text{th})^2} + 
	\frac{(\log ft^\text{th}_R(g_\text{A}, g_\text{pp}) - \log ft^\text{exp}_R)^2}
	{(\delta\log ft^\text{exp}_R)^2 + (\delta\log ft^\text{th})^2},
\end{align}
resulting in a two-dimensional parameter space that can be easily visualized. 
Fig.~\ref{fig:chi2} shows the $\chi^2$ fit based on Eq.~\eqref{eq:chi2_triplet_simple} 
and the individual contributions from the measurement of the left-leg (red) and 
right-leg (blue) beta/EC decays. In addition to the experimental errors, the $\chi^2$ 
fit includes a common theoretical uncertainty of $\delta\log ft^\text{th} = 10$\%. It 
has been chosen to be rather unrealistically small to show the effect of the experimental 
errors that otherwise are usually small compared to the model uncertainty. In the 
first three cases $(A,Z_0) = (100,42)$ (a), $(116,48)$ (b) and $(128,52)$ (c), the triplet includes a $2\nu\beta\beta$ decay isotope for which the 
half-life has been measured, $^{100}_{\ 42}$Mo, $^{116}_{\ 48}$Cd and $^{128}_{\ 52}$Te, respectively. The green 
band gives the correspondingly allowed $1\sigma$ parameter space. For $^{100}_{\ 42}$Mo 
(a) it overlaps well with the fit from the beta/EC decays. In the other two 
cases (b and c) there is a tension between single beta/EC decay and 
$2\nu\beta\beta$ decay data in the chosen model, but in both cases the discrepancy 
corresponds to a modest difference in $g_\text{A}$, by less than 30\%. 
We discuss this discrepancy and possible causes at the end of Sec.~\ref{sec:results_isobars}.
In the $A=128$ case (c) there are formally two best-fit solutions, the significance of which we  comment on below. The final scenario (d) illustrates a case where there is a tension between experimental data and theoretical predictions, i.e. the minimal $\chi^2$ is different from zero.

\begin{figure}[t!]
\centering
\subfloat[]{\includegraphics[width=0.49\columnwidth]{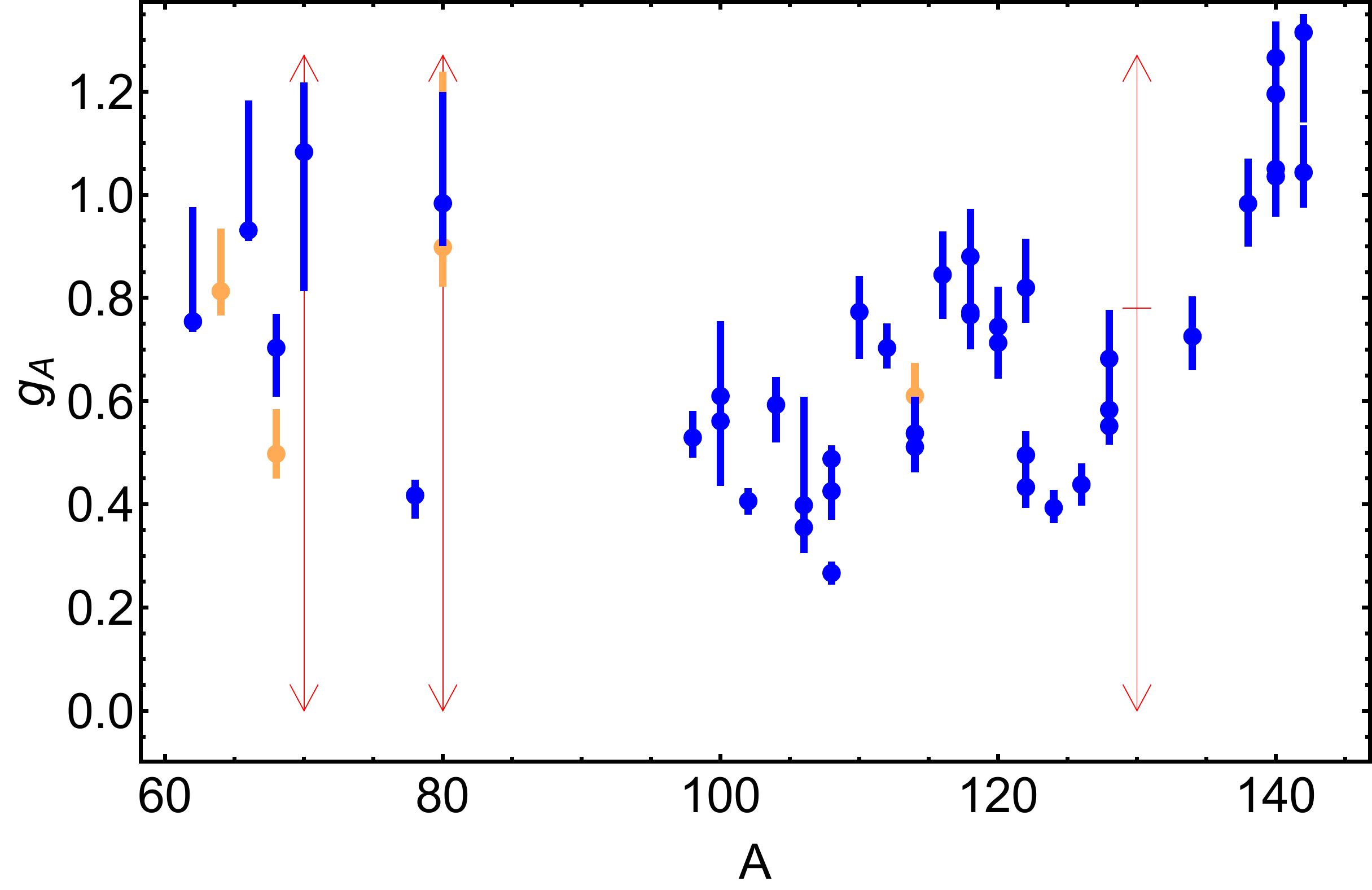}\label{fig:A_gA__A_gpp_a}}
\subfloat[]{\includegraphics[width=0.49\columnwidth]{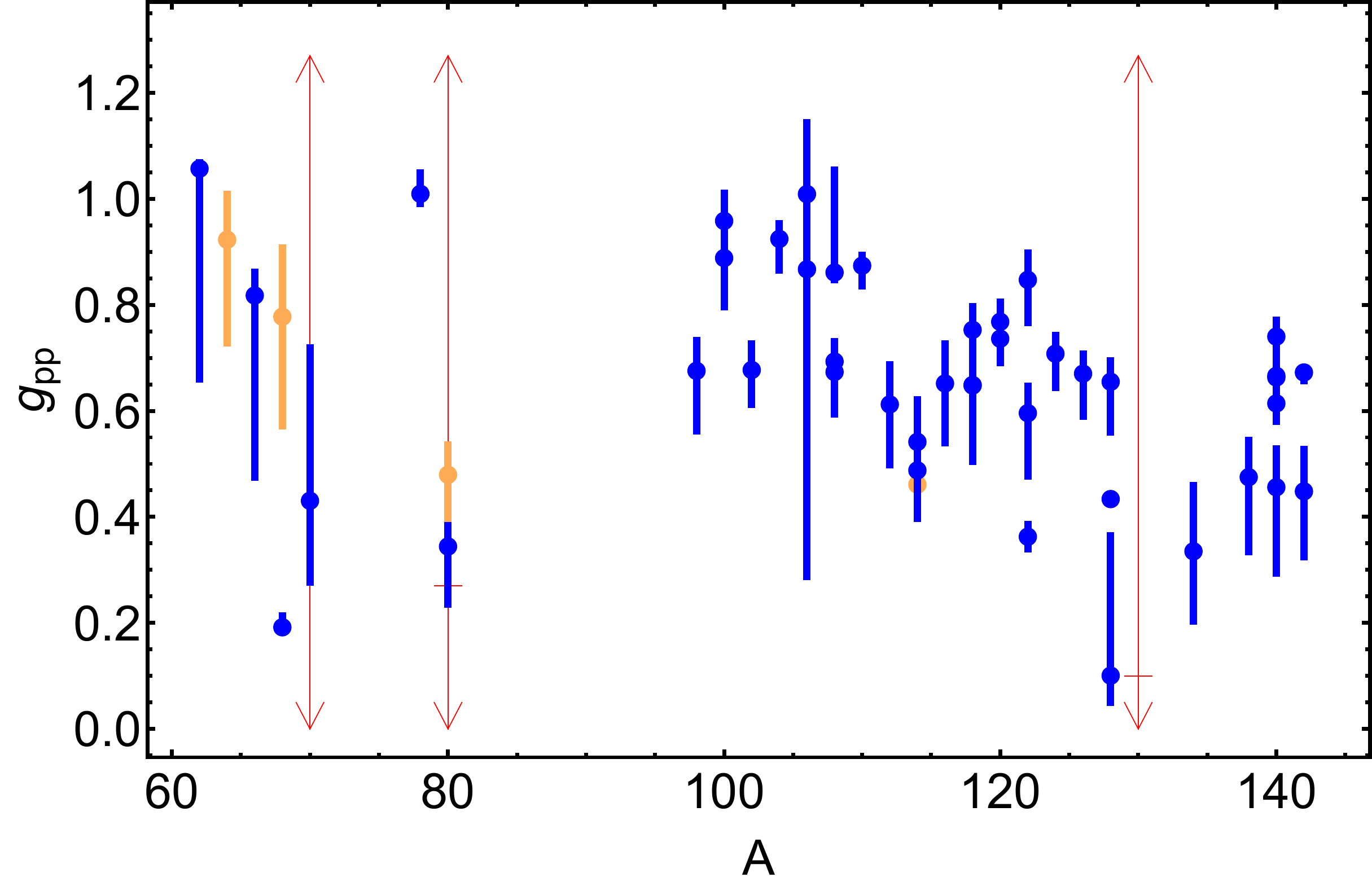}\label{fig:A_gA__A_gpp_b}}
\caption{$g_\text{A}$ (a) and $g_\text{pp}$ (b) determined in the individual triplet fits as a 
function of the mass number $A$, cf. Tabs.~\ref{tab:triplets}, \ref{tab:triplets2}. 
The error bars denote the $1\sigma$ parameter ranges. The dark blue and light orange values 
correspond to triplet fits with best fit solution $\chi^2_\text{min} = 0$ (within 
numerical tolerance) and $\approx 1$, respectively. The horizontal lines and associated vertical double arrows indicate fits with a strong tension with data where no meaningful error could be determined.}
\label{fig:A_gA__A_gpp}
\end{figure}
For the actual numerical determination of the best-fit parameters and their errors, 
we use Eq.~\eqref{eq:chi2_triplet}, where we include a 15\% uncertainty in the value(s) 
of the particle-hole nuisance parameter(s) $g_\text{ph}$ to model an additional theoretical 
uncertainty. Using the MCMC method described above, we determine the best-fit values 
and $1\sigma$ errors for $g_\text{pp}^\text{fit}$ and $g_\text{A}^\text{fit}$ in all 
triplets as given in Tabs.~\ref{tab:triplets} and \ref{tab:triplets2}. They are based 
on the fully marginalized distribution for the given parameter, but we omit secondary 
solutions for large $g_\text{pp}$ beyond the divergence, cf. 
Fig.~\ref{fig:chi2_c}. These large values of $g_\text{pp}$ make the pnQRPA
solutions unstable and in the worst case the whole set of pnQRPA solutions collapses
because the condition of small-amplitude motion of the RPA theory becomes seriously
violated \cite{Suhonen2007}.

In Tabs.~\ref{tab:triplets} and \ref{tab:triplets2}, the results highlighted with italic numbers correspond to fitting results with a minimal $\chi^2$ value of the order of $1$, 
indicating a slight tension between the experimental data and the theoretical 
predictions. As an example, the fitting of triplet $A = 68$, $Z_0 = 30$ is illustrated 
in Fig.~\ref{fig:chi2_d}. While the numerical fit in Table~\ref{tab:triplets} 
includes a larger theoretical uncertainty from the variation of the left- and 
right-leg particle-hole parameters $\gamma_\text{ph}^{L,R}$, this is not sufficient   
to achieve a vanishing $\chi^2_\text{min}$. Similar behavior occurs for the other triplets highlighted with italic numbers in the table. In such cases, the statistical uncertainty likely underestimates the true theoretical error. 

For the triplets $A = 80$, $Z_0 = 35$ and $A = 130$, $Z_0 = 54$, 
the minimal $\chi^2$ value is substantially different from zero ($\chi^2_\text{min} = 33$ 
and $15$, respectively), meaning that there is a large tension. We still give 
the nominal best fit values for $g_\text{A}$ and $g_{\rm pp}$, highlighted in bold in these 
cases, to indicate the tendency of the fit, but we do not quote an uncertainty. In fact, 
in both cases the best fit is achieved at the limit of the considered parameter space 
and the statistical uncertainty is rather meaningless. Finally, for $A = 70$, $Z_0 = 29$ 
no meaningful fit was achieved. 
In all other cases, a minimal $\chi^2 = 0$ (within numerical 
tolerance) was found. As can be seen in Tabs.~\ref{tab:triplets} and 
\ref{tab:triplets2}, problems to fit the experimental data mostly occur for lighter 
nuclei with $A \leq 80$. This could indicate the diminishing flexibility of the pnQRPA
model in going from the heavy nuclei, with large active single-particle model spaces, 
towards the lighter nuclei with small active model spaces, better suited for shell-model
description.

The results are also graphically illustrated in Fig.~\ref{fig:A_gA__A_gpp}, as a 
function of the mass number $A$. Analogous to the tables, the light orange points and $1\sigma$ error 
bars represent triplet fits with $\chi^2_\text{min} \approx 1$ whereas the blue points 
correspond to $\chi^2_\text{min} = 0$ (within numerical tolerance). 
The vertical double arrows (and horizontal lines for the nominal best fit, where applicable) indicate the cases with a strong 
tension with data 
as discussed above. As seen in the plots, the best fit values in these scenarios are 
still in the right ballpark of the neighboring fits. 

The strongest feature in the 
plots is the rise of $g_\text{A}$ with larger $A$ from $A = 98$ to 142 accompanied 
with a fall of $g_\text{pp}$. In this region the effective $g_\text{A}$ increases from a 
strongly quenched $g_\text{A} \approx 0.4$ around $A = 100$ to an essentially 
unquenched $g_\text{A} \approx 1.1$ around $A = 140$. Although there is a considerable 
spread in the values, the fitting results of triplets within the same mass number $A$ 
are largely compatible, illustrated by the overlapping error bars in 
Fig.~\ref{fig:A_gA__A_gpp}. The tendency for lighter nuclei ($A=80$ and below) is less 
clear and here the result is also affected by the large number of cases with tension to data. 
The tendency of a growing effective value of $g_\text{A}$ with mass number for the
$A\ge 100$ nuclei is in agreement with the linear model of Ref.~\cite{Pirinen2015}. From
Fig.~\ref{fig:A_gA__A_gpp_a} one can deduce the average value 
$g_\text{A}\approx 0.6$ for the $A\ge 100$ nuclei in accordance with the analysis
of Ref.~\cite{Pirinen2015}.

Using the thus fitted parameters, we calculate the predicted $2\nu\beta\beta$ decay 
half-lives for all relevant isotopes as listed in Table~\ref{tab:2vbbresults} under 
$[t_{1/2}^{(2\nu)}]_\text{triplet}$. The calculation includes the correlation among 
$g_\text{A}$, $g_\text{pp}$ and $g_\text{ph}$; i.e., we use the full probability 
density from the MCMC fit. Confirming the expectation of the simple two-dimensional 
fits shown in Fig.~\ref{fig:chi2}, the measured $2\nu\beta\beta$ decay half-life of 
$^{100}$Mo is consistent with the prediction within $1\sigma$. On the other hand, the 
predictions for $^{116}$Cd and $^{128}$Te are too small by a factor of about 2 compared 
to the experimental results. We further discuss this discrepancy at the end of the next section.
According to the calculated half-lives, the $2\nu\beta\beta$ decay of the nucleus 
$^{110}$Pd would be an interesting case to measure in the future.

\subsection{Fitting isobaric multiplets}
\label{sec:results_isobars}

%
\begin{figure}[t!]
\centering
\includegraphics[width=0.9\columnwidth]{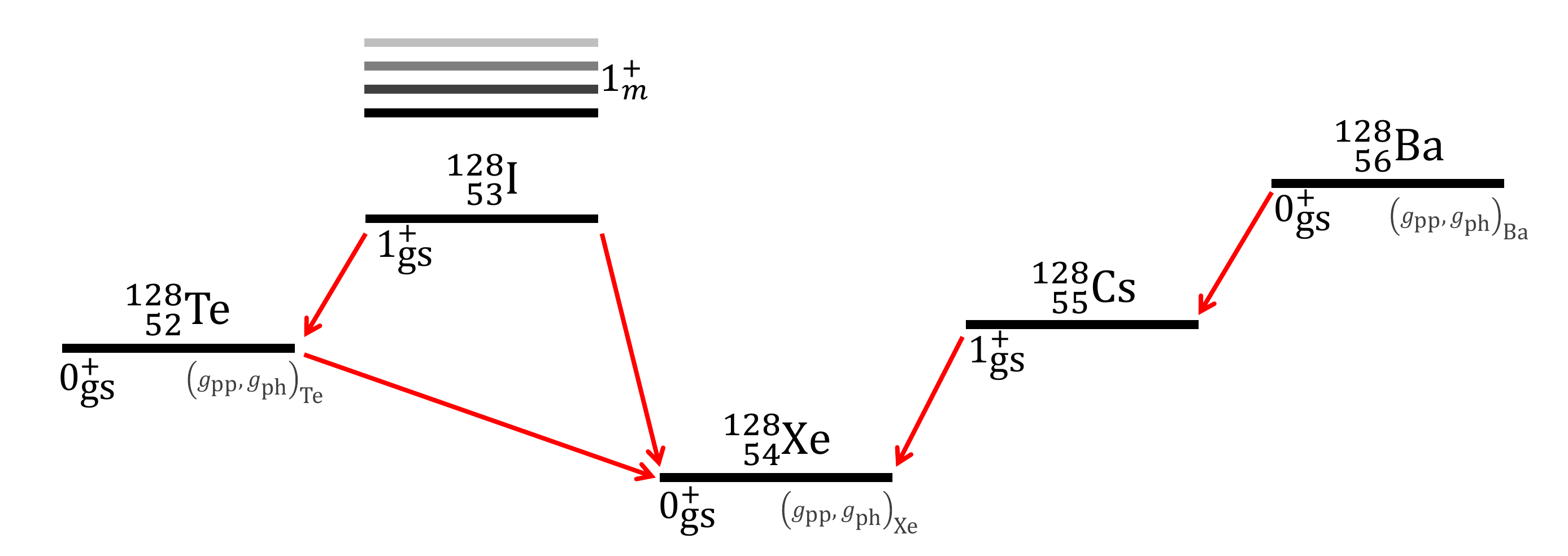}
\caption{Decay scheme for isotopes in the isobaric quintet $A=128$, $Z=52$-$56$. 
The different parameters $g_\text{pp}$ and $g_\text{ph}$ used in the fit for this multiplet 
are indicated for each even-even isotope in the system. Together with $g_\text{A}$, 
this leads to four (non-nuisance) parameters compared to four constraints from the beta/EC decays. The system also contains the $2\nu\beta\beta$ isotope $^{128}_{\ 52}$Te decaying to $^{128}_{\ 54}$Xe.}
\label{fig:schemeTe}
\end{figure}
Under the assumption that $g_\text{A}$ is a function of the mass number $A$ only, as 
justified in the Introduction, we can extend the analysis of individual triplets by 
fitting all beta/EC decays within a system (multiplet) of isobaric isotopes. This 
allows us to incorporate the full experimental information available. On the other
hand, we endeavor to include the full parametric uncertainty inherent in the 
pnQRPA models used in this study. This means that for each isobaric multiplet we
use a common $g_\text{A}$ and a set of ($g_\text{pp}$, $g_\text{ph}$) for each 
even-even isotope within the multiplet. The resulting multiplets are listed in 
Table~\ref{tab:isobarresults}, identified by $(A,Z_0)$ of the first isotope. The 
table indicates all isotopes in the multiplet by their atomic number and the arrows 
give the direction of the decays. The multiplicity within the isobaric systems ranges from three to seven isotopes. The 
triplets are obviously identical to the ones discussed in the previous section but we 
would like to note that we now allow for more parametric freedom because we use a 
separate $g_\text{pp}$ for each even-even isotope. Almost all triplets contain two 
even-even isotopes, i.e. two different $g_\text{pp}^L$ and $g_\text{pp}^R$ are used 
in the fit. Only the triplet $(98,39)$ contains a single even-even isotope and the fit 
performed is identical to the one in the previous section. 

In general, the degree 
of freedom in a fit is given by the number of free parameters (one $g_\text{A}$ and a $g_\text{pp}$ 
for each even-even isotope) minus the number of experimental constraints (number of 
measured beta/EC decays). This difference is given in Table~\ref{tab:isobarresults} 
under 'dof' indicating that we encounter underconstrained (dof $>0$), 
exactly constrained (dof $=0$) and overconstrained (dof $<0$) systems. While we also 
include a $g_\text{ph}$ for each even-even isotope, this does not change the number of 
degrees of freedom because the $g_\text{ph}$ are treated as nuisance parameters. An example 
of the decays and parameters involved is shown in Fig.~\ref{fig:schemeTe} in the case 
of the multiplet $(128,52)$.

\begin{table} 
\centering
\begin{tabular}{rr|l|c|l}
\hline
$A$ & $Z_0$ & \multicolumn{1}{|c|}{Multiplet} & dof & $g_\text{A}^\text{fit}$ \\
\hline
62	&	28	&	$28 \gets 29 \gets 30$                                     
& 3-2 & $	0.80	^{+	0.43	}_{-	0.01	}$ \\
64	&	28	&	$28 \gets 29 \to   30$                                     
& 3-2 & $	0.90	^{+	0.11	}_{-	0.09	}$\\
66	&	28	&	$28 \to   29 \to   30$                                     
& 3-2 & $	1.00	^{+	0.19	}_{-	0.16	}$\\
68	&	29	&	$29 \to   30 \gets 31 \gets 32$                            
& 3-3 & $	0.65	^{+	0.06	}_{-	0.07	}$\\
70	&	29	&	$29 \to \underline{30} \gets 31 \to   32$                  
& 3-3 & {\textbf -} \\
78	&	34	&	$34 \gets 35 \to   36$                                     
& 3-2 & $	0.35	^{+	0.59	}_{-	0.02	}$\\
80	&	33	&	$33 \to \underline{34} \gets 35 \to   36 \gets 37$         
& 3-4 & {\textbf{1.40}}\\
98	&	39	&	$39 \to   40 \to   41$                                     
& 2-2 & $	0.53	^{+	0.04	}_{-	0.03	}$\\
100	&	41	&	$41 \to \underline{42} \gets 43 \to   44$                  
& 3-3 & $	0.37	^{+	0.22	}_{-	0.00	}$\\
102	&	42	&	$42 \to   43 \to   44$                                     
& 3-2 & $	0.34	^{+	0.16	}_{-	0.00	}$\\
104	&	44	&	$\underline{44} \gets 45 \to   46$                         
& 3-2 & $	0.59	^{+	0.28	}_{-	0.10	}$\\
106	&	45	&	$45 \to   46 \gets 47 \to   48$                            
& 3-3 & $	0.40	^{+	0.02	}_{-	0.02	}$\\
108	&	44	&	$44 \to   45 \to   46 \gets 47 \to   48$                   
& 4-4 & $\mathit{0.41	^{+	0.01	}_{-	0.01	}}$\\
110	&	46	&	$\underline{46} \gets 47 \to   48$                         
& 3-2 & $	0.71	^{+	0.38	}_{-	0.13	}$\\
112	&	48	&	$48 \gets 49 \to   50$                                     
& 3-2 & $	0.67	^{+	0.19	}_{-	0.03	}$\\
114	&	46	&	$46 \to   47 \to \underline{48} \gets 49 \to   50$         
& 4-4 & $\mathit{0.60	^{+	0.03	}_{-	0.03	}}$\\
116	&	48	&	$\underline{48} \gets 49 \to   50$                         
& 3-2 & $	0.68	^{+	0.38	}_{-	0.01	}$\\
118	&	48	&	$48 \to   49 \to   50 \gets 51 \gets 52$                   
& 4-4 & $	0.75	^{+	0.06	}_{-	0.04	}$\\
120	&	48	&	$48 \to   49 \to   50 \gets 51$                            
& 3-3 & $	0.71	^{+	0.06	}_{-	0.05	}$\\
122	&	48	&	$48 \to   49 \to   \underline{50}\,\,\,\,|\,\,\,52 
\gets 53 \gets 54 \gets 55$	 & 5-5 &
$\mathit{0.49	^{+	0.03	}_{-	0.03	}}$\\
124	&	54	&	$54 \gets 55 \gets 56$                                     
& 3-2 & $	0.34	^{+	0.20	}_{-	0.02	}$\\
126	&	54	&	$54 \gets 55 \gets 56$                                     
& 3-2 & 
$	0.35	^{+	0.20	}_{-	0.02	}$\\
128	&	52	&	$\underline{52} \gets 53 \to   54 \gets 55 \gets 56$       
& 4-4 & 
$\mathit{0.59	^{+	0.05	}_{-	0.06	}}$\\
130	&	54	&	$54 \gets 55 \to   56$                                     
& 3-2 & {\textbf{0.78}}\\
134	&	56	&	$56 \gets 57 \gets 58$                                     
& 3-2 & $	0.72	^{+	0.10	}_{-	0.08	}$\\
138	&	58	&	$58 \gets 59 \gets 60$                                     
& 3-2 & $	0.92	^{+	0.20	}_{-	0.09	}$\\
140	&	58	&	$58 \gets 59 \gets 60 \gets 61 \gets 62 \gets 63 \gets 64$ 
& 5-6 & $	1.10	^{+	0.07	}_{-	0.09	}$\\
142	&	60	&	$60 \gets 61 \gets 62 \gets 63$                            
& 3-3 & $	1.20	^{+	0.07	}_{-	0.12	}$\\
\hline
\end{tabular}
\caption{Isobaric multiplets of $\beta^+$/EC and $\beta^-$ decaying isotopes studied 
in the present paper. An isobaric multiplet is identified by the mass number $A$ and 
the lowest atomic number $Z_0$ among its isotopes. The individual isotopes are indicated 
with their atomic number and the arrows denote the direction of the relevant 
$\beta^+$/EC decays. The column 'dof' gives the degrees of freedom, i.e. the number of 
free parameters (one $g_\text{A}$ and a $g_\text{pp}$ for each even-even isotope) minus 
the number of experimental constraints. $2\nu\beta\beta$ decaying isotopes are underlined. 
The values of $g_\text{A}$ are determined in the isobar fits described in 
Sec.~\ref{sec:results_isobars}. Cases in which the best fit $\chi^2_\text{min}$ is in 
the range $[0.8, 1.5]$, indicating slight tension with data, are highlighted with italic numbers. Cases with stronger tension are highlighted in bold. In all other 
cases a $\chi^2_\text{min}$ of the order expected by the degrees of freedom was found.}
\label{tab:isobarresults} 
\end{table}
\begin{figure}[t!]
\centering
\subfloat[]{\includegraphics[width=0.32\columnwidth]{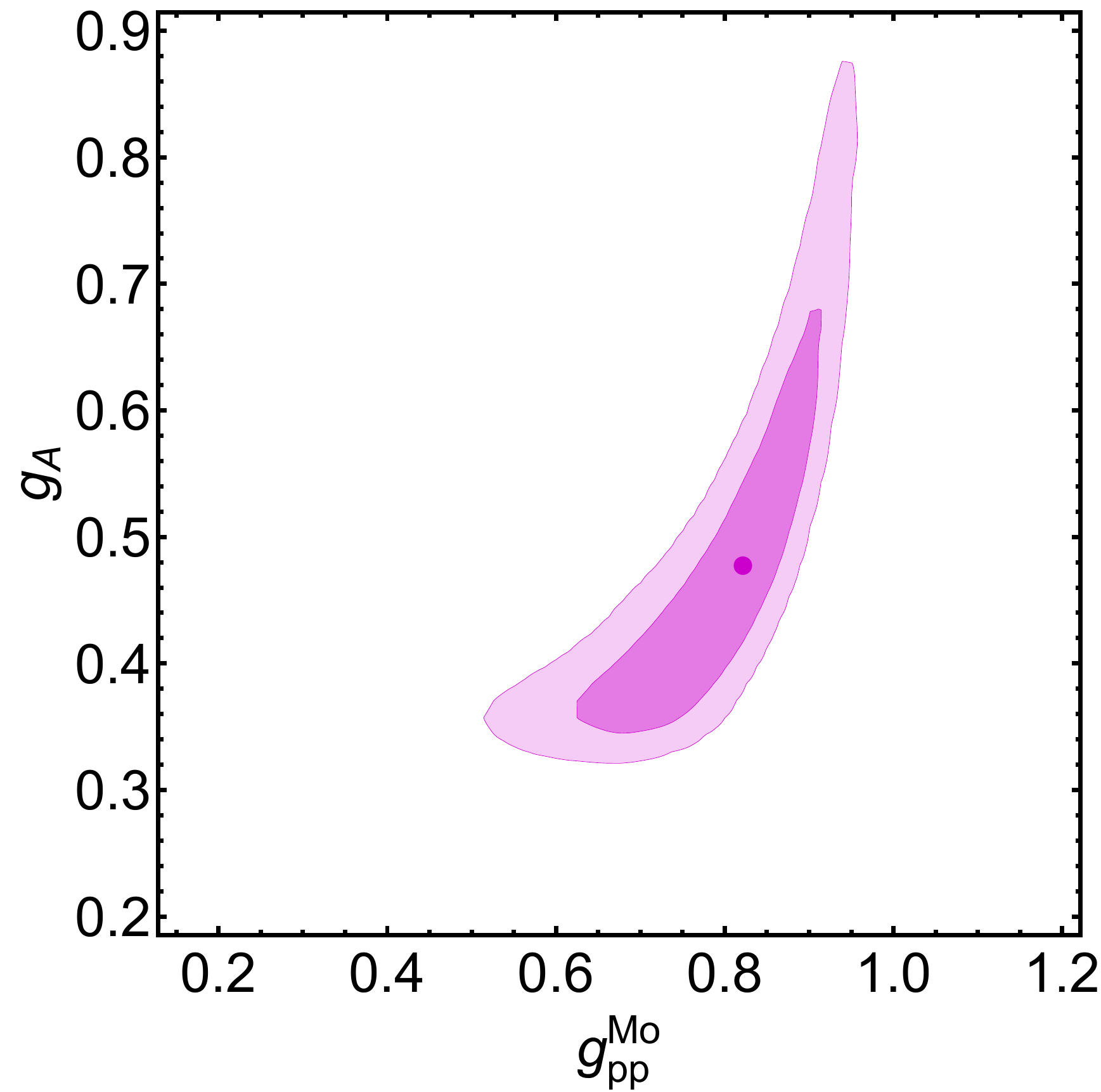}\label{fig:likelihoods_isobar_a}}
\subfloat[]{\includegraphics[width=0.32\columnwidth]{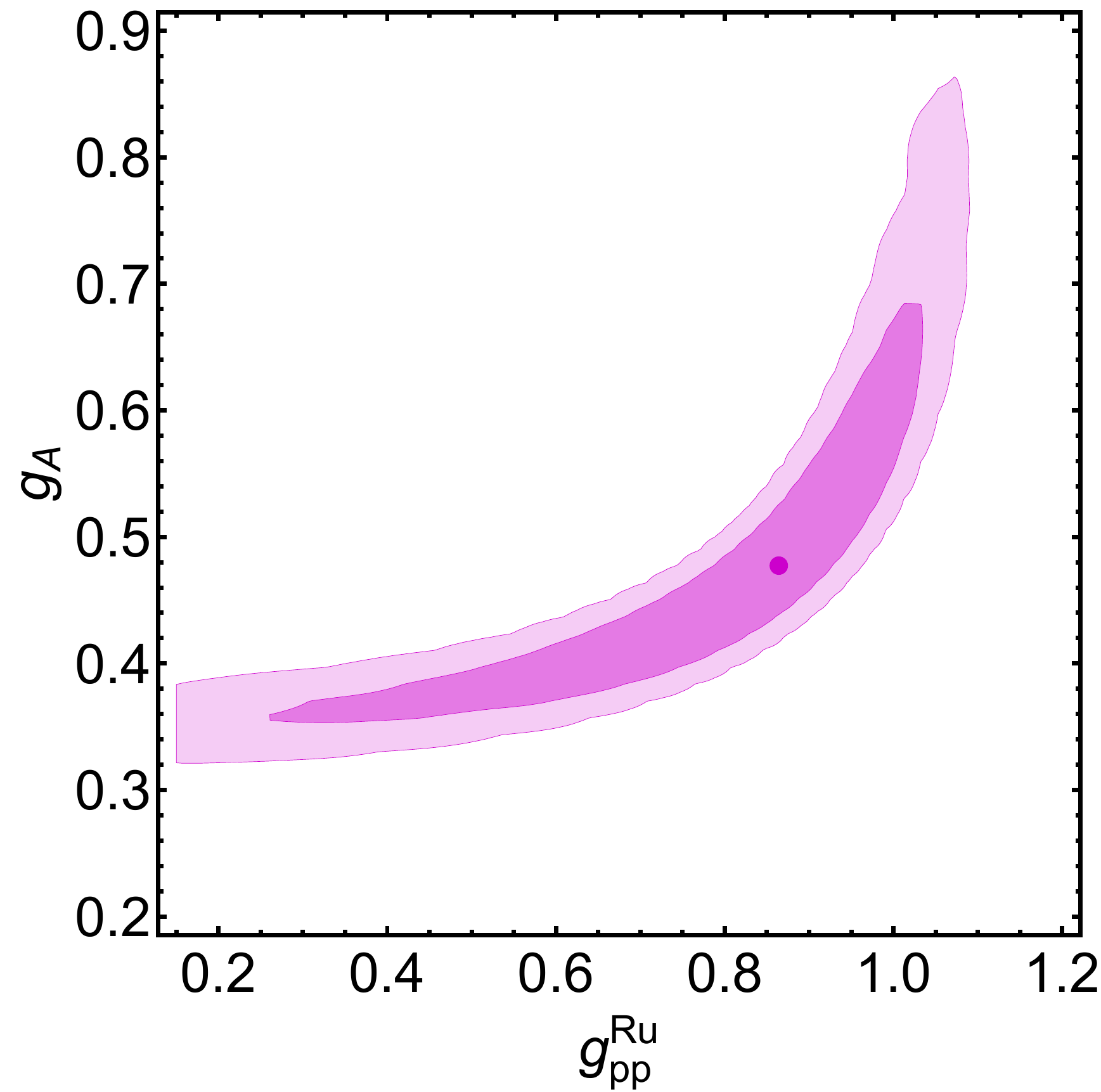}\label{fig:likelihoods_isobar_b}}
\subfloat[]{\includegraphics[width=0.32\columnwidth]{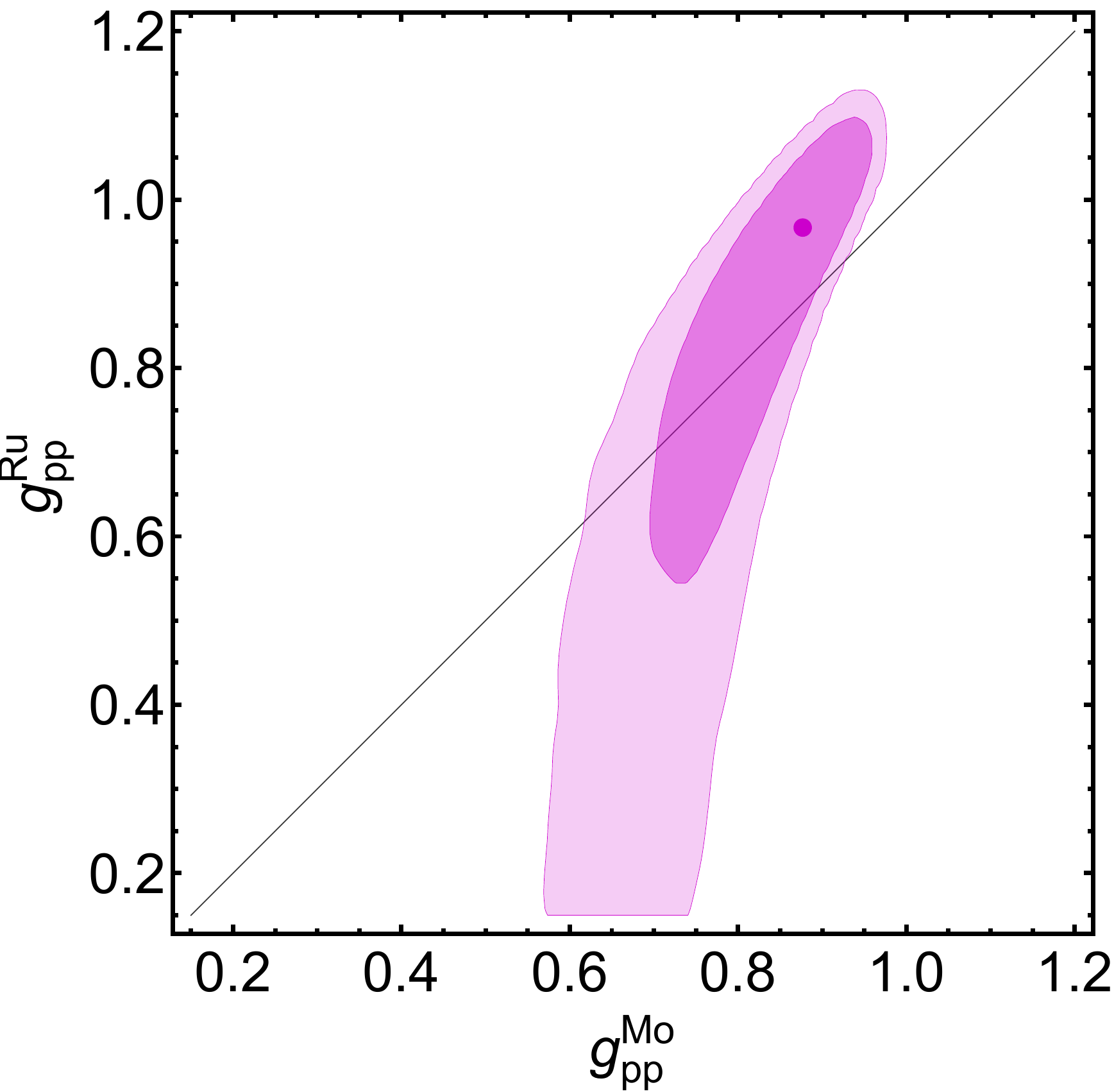}\label{fig:likelihoods_isobar_c}}
\caption{Probability density function for the fit of the isobaric system $A=100$, $Z_0=41$ 
in three marginalizations of the parameter space: $g_\text{A} - g_\text{pp}^\text{Mo}$ 
(a), $g_\text{A} - g_\text{pp}^\text{Ru}$ (b) and 
$g_\text{pp}^\text{Mo} - g_\text{pp}^\text{Ru}$ (c). The colored areas denote the 
$1\sigma$ and $2\sigma$ extent of the parameter space and the dot denotes the maximum.}
\label{fig:likelihoods_isobar}
\end{figure}
The MCMC fitting procedure is analogous to the triplet case with a $\chi^2$ function of 
the form of Eq.~\eqref{eq:chi2_triplet}: there is a contribution for each observable 
(beta/EC decay), with the theoretically calculated $\log ft$ values depending on the 
appropriate $g_\text{pp}^i$ and $\gamma_\text{ph}^i$ and the global $g_\text{A}$. In addition, 
there is a nuisance term for each of the particle-hole parameters $\gamma_\text{ph}^i$ 
where we again use an uncertainty of $\delta\gamma_\text{ph}^i = 0.15$. As before, we 
use the experimental measurements and errors of the beta/EC decay $\log ft$ values 
from Tabs.~\ref{tab:triplets} and \ref{tab:triplets2}. 

An example of the result of an isobar fit is shown in Fig.~\ref{fig:likelihoods_isobar} displaying 
the probability density functions in three different marginalized parameter planes as 
derived in the isobaric system $(100,41)$ containing the $2\nu\beta\beta$ decay 
isotope $^{100}_{\ 42}$Mo. It is a quartet with two even-even isotopes and thus described 
by the following three parameters, $g_\text{A}$, $g_\text{pp}^\text{Mo}$ and $g_\text{pp}^\text{Ru}$, and two 
nuisance parameters, $\gamma_\text{ph}^\text{Mo}$ and $\gamma_\text{ph}^\text{Ru}$. The system is exactly constrained and 
consistent; thus the minimal $\chi^2$ is zero. Nevertheless, there is a sizable 
uncertainty and especially the parameter $g_\text{pp}^\text{Ru}$ can vary strongly and 
no lower limit can be determined at $2\sigma$ within the chosen parameter range. In this 
case the two particle-particle parameters can be very different. For example, the 
combination $g_\text{pp}^\text{Mo} \approx 0.7$ and $g_\text{pp}^\text{Ru} \approx 0.1$ is 
allowed by the combined experimental data within $2\sigma$. Another consequence of this 
is that $g_\text{A}$ is effectively suppressed compared to the triplet fit because lower values are statistically preferred for small $g_\text{pp}^\text{Ru}$. The experimental $\log ft$ 
errors and the variation of $\gamma_\text{ph}^\text{Mo}$ and $\gamma_\text{ph}^\text{Ru}$ 
provide an additional source of uncertainty.

The results of the fits are displayed in Table~\ref{tab:isobarresults}, which lists the 
best-fit values and $1\sigma$ uncertainties of $g_\text{A}$ for all multiplets. As in 
the case of the individual triplets, italicized values denote multiplets in which 
there is a slight tension between the combined experimental data and the 
theoretical predictions. Comparing with Tabs.~\ref{tab:triplets}, 
\ref{tab:triplets2}, the tension for $A = 62,80$ is resolved by 
introducing the additional freedom in the multiplet fits. On the other hand, the 
multiplets $A=108, 114, 122, 128$ exhibit slight tension, due to the larger number 
of simultaneous constraints that dominate over the additional freedom. The strong 
tension for $A=70, 80, 130$ remains in the multiplet approach.

\begin{figure}[t!]
\centering
\subfloat[]{\includegraphics[width=0.49\columnwidth]{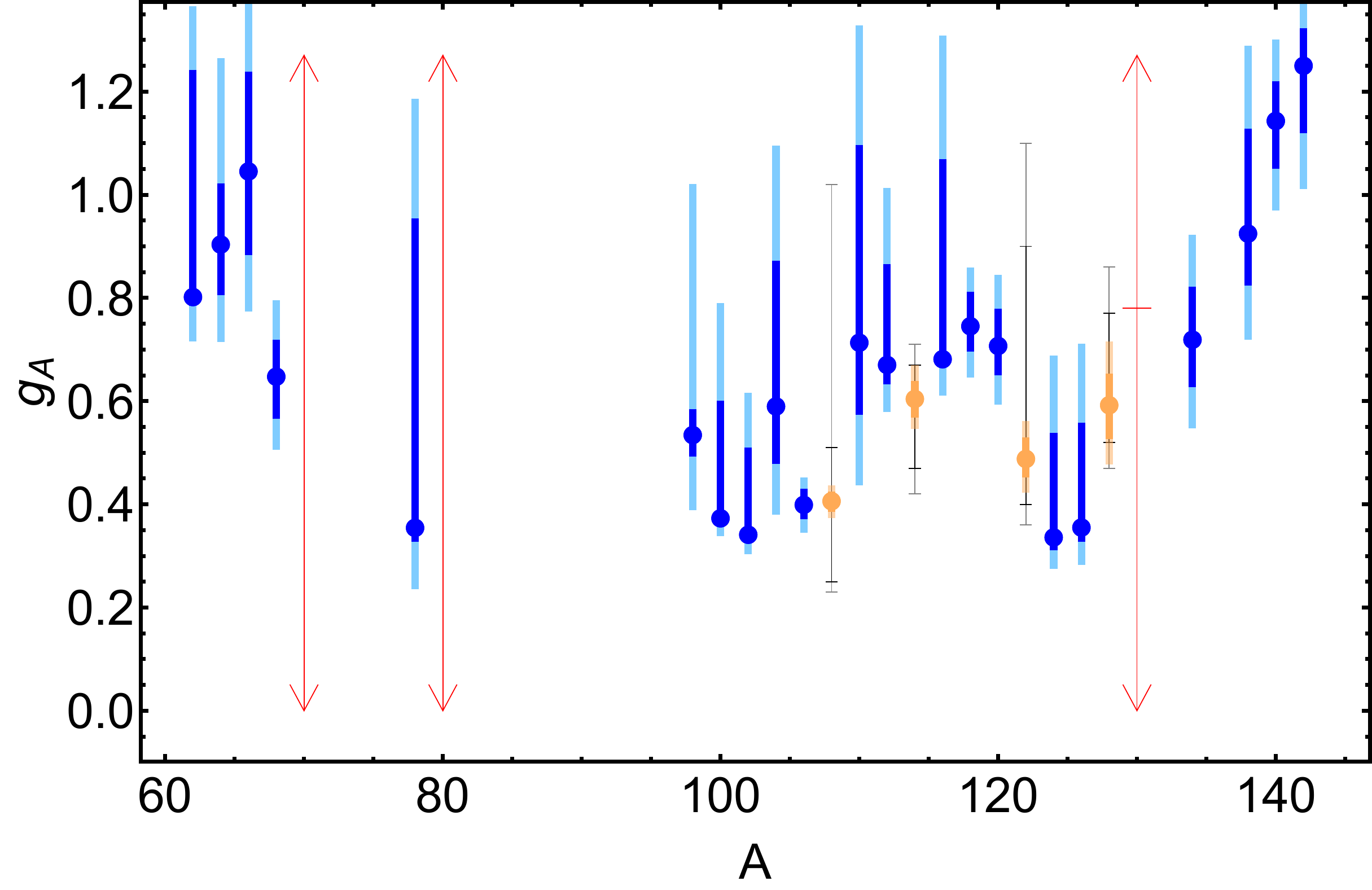}\label{fig:A_gA__A_gpp_isobar_a}}
\subfloat[]{\includegraphics[width=0.49\columnwidth]{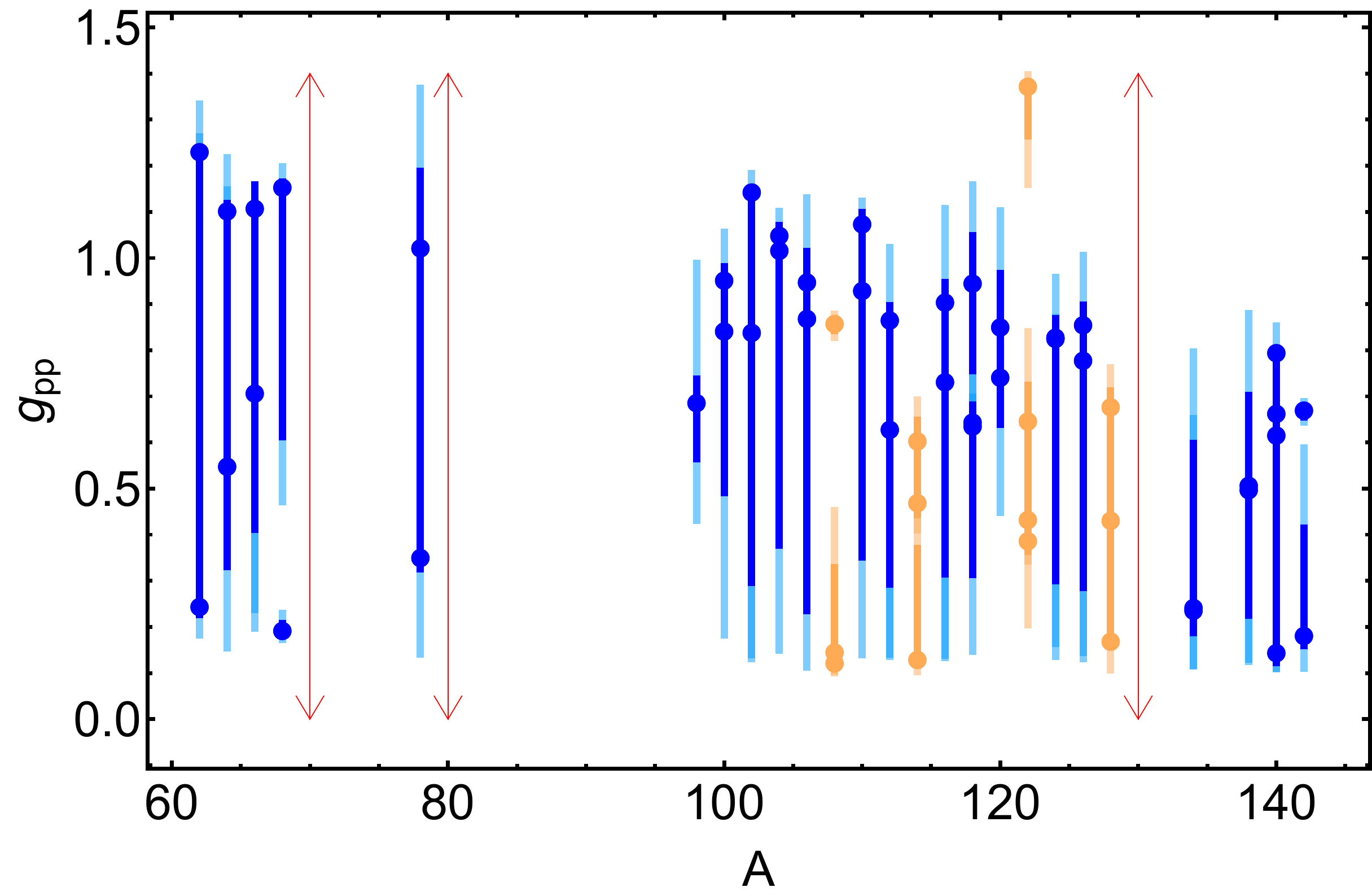}\label{fig:A_gA__A_gpp_isobar_b}}
\caption{$g_\text{A}$ (a) and $g_\text{pp}$ (b) determined in the isobaric multiplet fits as a function of the 
mass number $A$, cf. Table~\ref{tab:isobarresults}. The error bars denote the 
$1\sigma$ and $2\sigma$ parameter ranges. The blue values correspond to multiplets 
with a best fit solution as expected by the degrees of freedom whereas light orange 
values denote a slight tension with $\chi^2_\text{min} \approx 1$. In these cases, 
the underlying black and gray error lines give the $1\sigma$ and $2\sigma$ range among 
the triplet fits, respectively, for comparison. The horizontal lines and associated vertical double arrows indicate fits with a strong tension where no meaningful error 
could be determined.}
\label{fig:A_gA__A_gpp_isobar}
\end{figure}

The results are also graphically displayed in Fig.~\ref{fig:A_gA__A_gpp_isobar_a}, where 
we plot the extracted values of $g_\text{A}$ as a function of the mass number $A$, also 
showing the $1\sigma$ and $2\sigma$ uncertainties. It can be directly compared to 
Fig.~\ref{fig:A_gA__A_gpp_a}, and it can be seen that the general behavior 
exhibited is similar in the two plots. Due to the combined fit of all isobaric nuclei, 
the dependence on $A$ is smoother in Fig.~\ref{fig:A_gA__A_gpp_isobar_a}, albeit with 
sometimes considerable uncertainty for some multiplets, suggesting a systematic 
dependence of $g_\text{A}$ with the mass number $A$ within the range $98 \leq A \leq 142$. 
The multiplets $A = 108, 114, 122, 128$ with slight tension correspond to 
triplets with comparatively large differences in the fitted $g_\text{A}$ values. This 
tension between the triplet $g_\text{A}$ values leads to a worse fit when combining 
within a multiplet and the resulting uncertainty likely underestimates the true error 
in $g_\text{A}$. For comparison in these cases, Fig.~\ref{fig:A_gA__A_gpp_isobar_a} also 
shows the $1\sigma$ and $2\sigma$ range derived in the associated triplet fits (i.e. 
the minimal and maximal extent in $g_\text{A}$ among all triplets at the given 
uncertainty). This likely provides a more realistic estimate of the 
uncertainty in $g_\text{A}$.

Fig.~\ref{fig:A_gA__A_gpp_isobar_b} shows $g_\text{pp}$ as a function of $A$ within the isobaric fits where each dot represents the best fit point of an individual $g_\text{pp}^i$. As is illustrated in Fig.~\ref{fig:likelihoods_isobar} and discussed above, the freedom of allowing a different $g_\text{pp}$ per even-even isotope introduces a degeneracy where individual $g_\text{pp}$ can vary strongly. Fig.~\ref{fig:A_gA__A_gpp_isobar_b} demonstrates that in almost all cases at least one of the $g_\text{pp}$ can be small, hitting the limit of the chosen variable range. As a result the apparent spread among the $g_\text{pp}$ is large and there is only a tendency of squeezing the range to lower values for increasing $A$. It should be kept in mind that the fitting procedure also means that the individual $g_\text{pp}$ values are correlated to satisfy the experimental constraints. Despite the large freedom and large variability of the $g_\text{pp}^i$ it is remarkable that the corresponding uncertainty of $g_\text{A}$ remains comparatively small. This could indicate a robustness of the underlying pnQRPA-based nuclear-structure calculations against parameter variations.

Using the thus fitted parameters, we calculate the predicted $2\nu\beta\beta$ decay 
half-lives for all relevant isotopes as listed in Table~\ref{tab:2vbbresults} under 
$[t_{1/2}^{(2\nu)}]_\text{multiplet}$. The calculation includes the correlation among 
the parameters $g_\text{A}$, $g_\text{pp}^i$, and $\gamma_\text{ph}^i$; i.e., we use the 
full probability density from the MCMC fit. Unfortunately, many of the $2\nu\beta\beta$ 
isotopes are part of multiplets with a slight or severe tension between experimental 
data and theoretical prediction of the EC/beta decay fit. These are highlighted in 
Table~\ref{tab:2vbbresults} as usual. The uncertainties quoted in these cases are likely
underestimates. The multiplet fits incorparating $^{116}$Cd and $^{128}$Te are not 
able to relax the tension with the measured $2\nu\beta\beta$ decay half-lives in 
these isotopes.

On the other hand, in the analysis carried out in Ref.~\cite{Barea2013} a very weakly decreasing $A$ dependence was obtained for the effective value of $g_\text{A}$ in the IBA-2 and ISM frameworks. The result was based on the comparison of the computed and experimental half-lives of the $2\nu\beta\beta$ decays. This weak trend is in line with the analysis of Ref.~\cite{Pirinen2015} where it was observed that the constant $g_\text{A}=0.6$ reproduces better the measured $2\nu\beta\beta$ half-lives than the growing trend for $g_\text{A}$ obtained by the $\beta$-decay analysis (see Table~VI of Ref.~\cite{Pirinen2015}). In fact, having a look at Fig.~\ref{fig:chi2} as well as Table~\ref{tab:2vbbresults} of this work, in both the triplet and multiplet analyses a decreasing multiplicative factor of $g_\text{A}$, as a function of the mass number $A$, would bring the computed half-lives closer to the measured ones. The fact that the analyses of the $\beta$ decays and the $2\nu\beta\beta$ decays give opposite trends is a sign that there are basic differences between the two modes of decay. One difference is that for the $2\nu\beta\beta$ decays more than one low-lying state can contribute and the contribution coming from the first virtual $1^+$ state of the intermediate nucleus does not exhaust the whole $2\nu\beta\beta$ NME (see Table~V of Ref.~\cite{Pirinen2015}). The balance between the first contribution and the rest depends on the value 
of $g_\text{pp}$ in a pnQRPA calculation, and the correlations of $g_\text{A}$ and $g_\text{pp}$ are most likely different for the $\beta$ and $2\nu\beta\beta$ decays.


\section{Conclusions}
\label{sec:conclusions}

In this work we studied single $\beta^+$/EC and $\beta^-$ decays for a number of 
isobaric triplets and more complicated isobaric chains of nuclei in the framework of 
the proton-neutron quasiparticle random-phase approximation. 

The present calculations have been done  with $G$-matrix-based two-nucleon interactions. 
By letting the value of the axial-vector coupling constant $g_\text{A}$ vary freely, 
together with the particle-particle interaction strength parameter $g_\text{pp}$, we 
have performed MCMC-based statistical analyses to chart 
the effective values of $g_\text{A}$ in pnQRPA-based nuclear-structure calculations. 
Within the statistical fits of complete isobaric chains of nuclei we incorporated full 
parametric uncertainty of the nuclear model by using independent $g_\text{pp}$ per 
even-even isotope as well as an uncertainty in the particle-hole parameter $g_\text{ph}$. 
We thus not only confirm previous results of an apparent quenching of $g_\text{A}$ 
in an extended analysis but we also provide a realistic estimate of the parametric 
uncertainty inherent in the nuclear model. This is important, also to compare with 
other theory frameworks.

These findings may have some bearing on the studies of the contributions of the low-lying $1^+$ intermediate states to the highly interesting $0\nu\beta\beta$ NMEs. The relation of our present results to the values of the $0\nu\beta\beta$ NMEs remains still an open issue but we view the present study as an incentive to tackle these issues in future investigations. 


\section*{Acknowledgements}
This work was supported by the Academy of Finland under the Finnish 
Center of Excellence Program 2012-2017 (Nuclear and Accelerator Based 
Program at JYFL) and by the U. K. Royal Society International Exchanges Grant IE130084. 

\section*{References}

\end{document}